\documentclass[%
 aip,
 amsmath,amssymb,
 reprint,%
]{revtex4-1}

\usepackage{graphicx}
\usepackage[utf8]{inputenc}
\usepackage[T1]{fontenc}
\usepackage{etoolbox}
\usepackage{hyperref}
\usepackage{multirow}
\usepackage{booktabs}
\usepackage{xcolor}
\usepackage{flafter}   
\usepackage{titlesec}
\renewcommand{\thesubsubsection}{\roman{subsubsection}}
\newcommand{\colonaftersub}[1]{#1:}
\titleformat{\subsubsection}[runin]{\normalfont\bfseries}{\thesubsubsection}{1em}{\colonaftersub}

\makeatletter
\def\@email#1#2{%
 \endgroup
 \patchcmd{\titleblock@produce}
  {\frontmatter@RRAPformat}
  {\frontmatter@RRAPformat{\produce@RRAP{*#1\href{mailto:#2}{#2}}}\frontmatter@RRAPformat}
  {}{}
}%
\makeatother

\newcommand{\Jb}{J_B}
\newcommand{\Joos}{J_{\mathrm{S}}}
\newcommand{\Jnorm}{J_{\mathrm{norm}}}
\newcommand{\units}{\mathrm{T}^2\mathrm{m}^2}
\newcommand{\xzero}{x_0}
\newcommand{\xpert}{\tilde{x}_0}
\newcommand{\sigt}{\sigma_{\mathrm{T}}}
\newcommand{\sigs}{\sigma_{\mathrm{S}}}

\begin{document}

\title{An In-depth Study of Stochastic Stellarator Coil Optimization}

\author{Vincenzo Gallegos}
\affiliation{\footnotesize{Department of Applied Physics, Tandon School of Engineering, New York University, New York, NY,~USA}\looseness=-1}
\email{vmg6966@nyu.edu}

\author{Evan Laljit}

\affiliation{\scriptsize{Department of Physics, College of Arts and Sciences, New York University, New York, NY,~USA}\looseness=-1}
\affiliation{\scriptsize{Department of Applied Physics and Applied Mathematics, Columbia University, New York, NY,~USA}\looseness=-1}

\author{Julianne Stratton}
\affiliation{\scriptsize{Courant Institute School of Mathematics, Computing, and Data Science, New York University, New York, NY,~USA}\looseness=-1}

\author{Gabe Perez-Giz}
\affiliation{\footnotesize{Department of Applied Physics, Tandon School of Engineering, New York University, New York, NY,~USA}\looseness=-1}

\author{Alan Kaptanoglu}
\affiliation{\scriptsize{Courant Institute School of Mathematics, Computing, and Data Science, New York University, New York, NY,~USA}\looseness=-1}
\affiliation{\scriptsize{William E. Boeing Department of Aeronautics and Astronautics, University of Washington, Seattle, Washington,~USA}\looseness=-1}


\begin{abstract}
We extend stochastic coil optimization for stellarators and analyze the effect of optimizing for coil manufacturing errors, assembly errors represented by centroid translations and angular offsets, and errors in the coil currents. We further analyze the impact of different perturbation methods and use them in conjunction with the Augmented Lagrangian algorithm. For three stellarators, we show that stochastic methods regularize the coil optimization landscape and improve robustness in every case, with the magnitude of the improvement varying substantially across different plasmas. Our metric for coil robustness, $Q$, is improved by $4.50\%$ to $729\%$ across different configurations. Lastly, we perform numerous hyperparameter scans, including e.g. scanning the training perturbation size and the out-of-sample (OOS) evaluation perturbation size and find coupled and de-coupled training tolerance gains ranging from $0.09$\,mm to $1.35$\,mm depending strongly on the particular stellarator.
\end{abstract}

\maketitle

\section{Introduction}
\label{sec:Introduction}

Stellarators are a type of device designed to magnetically confine high-temperature plasmas within a toroidal volume to achieve sustained nuclear fusion. Tokamaks are axisymmetric, whereas stellarators can possess more general three-dimensional symmetries such as quasisymmetry~\cite{boozer_transport_1983}. The magnetic-field requirements of quasisymmetry are found by shape optimization of the plasma boundary (``stage-one'' optimization). The confining magnetic field is typically generated by external coils that satisfy the boundary condition on the plasma boundary, $B\cdot\hat{n} = 0$, requiring another optimization problem to determine the coils (``stage-two'' optimization). The complex shape of the three-dimensional magnetic field typically leads to complex coils. This introduces engineering challenges and increases costs from manufacturing, assembly, and sensitivity issues.
Errors or changes in the geometry of the coils can cause large deviations from the desired plasma equilibrium, degrading fusion conditions~\cite{andreeva_analysis_2004,andreeva_influence_2017,brooks_coil_2003,cao_error_2026,hammond_experimental_2016,liu_influence_2026}.

Engineering constraints have been shown to correlate directly with cost. An investigative report from Princeton Plasma Physics Laboratory explains that the estimated cost and schedule for NCSX grew as its technical requirements and risks became better understood, contributing to cancellation in 2008~\cite{neilson_lessons_2010}. NCSX adopted a 1.5\,mm modular-coil tolerance after Monte Carlo error-field studies evaluated resonant islands and the usable magnetic-flux volume~\cite{brooks_coil_2003}. It contributed to CNC reprogramming~\cite{heitzenroeder_production_2005}, redesigns of coil interfaces~\cite{freudenberg_performance_2009}, and difficult fabrication~\cite{dudek_ncsx_2007}.

The tolerance numbers quoted for different devices typically have different sources. W7-X specified a 2\,mm construction-accuracy limit for its superconducting winding packs, motivated by control of resonant magnetic-field errors and the associated island structure~\cite{rummel_accuracy_2004,andreeva_analysis_2004,kislinger_correction_2005}. LHD required helical-coil manufacturing within $\pm2$\,mm to preserve a fine magnetic surface while maintaining rigidity under electromagnetic loading~\cite{imagawa_design_1998}. For CFQS, Shimizu \textit{et al.} found that special stellarator-symmetric radial, vertical, and tilt errors below approximately 10\,mm had little effect on magnetic surfaces, rotational transform, magnetic well, or effective helical ripple~\cite{shimizu_consideration_2019}. That result is a physics-based sensitivity range for a restricted error class, not an as-built tolerance. Tabletop stellarators provide an additional route for experimentally studying such effects~\cite{gil_manufacturing_2025}.

Deterministic stage-two coil optimization is well established but often produces coil--plasma designs that are sensitive to manufacturing errors~\cite{fu_global_2025,giuliani_direct_2024,mcgreivy_optimized_2021,paul_adjoint_2018,wiedman_coil_2024}. Brooks and Reiersen's NCSX study evaluated random coil errors through the toroidal-flux fraction occupied by islands~\cite{brooks_coil_2003}. Even in the 1.5\,mm ensemble, worst cases produced significant island-fractions, which motivated the tighter project specification. 

Manufacturing and assembly errors were major obstacles for NCSX, and deformation under electromagnetic loads provides an additional error source~\cite{heinemann_design_2005}. W7-X structural analyses and equilibrium studies explicitly include gravity, cool-down, electromagnetic loading, and assembly nonconformities rather than treating all deviations as a single manufacturing tolerance~\cite{bykov_specific_2014,andreeva_influence_2017}. On CNT, photogrammetry found a largest local marker departure of $44\pm3$\,mm, whereas best-fit rigid translations of the two interlocked coils were 5 and 22\,mm with tilts of $1.0^\circ$ and $1.3^\circ$~\cite{hammond_experimental_2016}. The Stellaris concept study reports a modeled peak structural displacement of 15\,mm under design loading, not an as-built measurement~\cite{lion_stellaris_2025}. Error fields can be mitigated with correction coils~\cite{kalish_ncsx_2009,risse_purpose_2011}, although these systems add cost and operational complexity.


Previous work has attempted to find more robust coil solutions by stochastic optimization techniques that take potential error sources into account. Wechsung \textit{et al.} performed simultaneous optimization of the plasma (although these calculations were performed in vacuum) and coils (``single-stage'' optimization) with a stochastic objective~\cite{wechsung_single-stage_2022}. Other work has studied stochastic optimization for global searches~\cite{glas_global_2022} and with winding-surface coils~\cite{lobsien_stellarator_2018}. Wechsung \textit{et al.} also investigated post-processing corrections in which measured coil errors were mitigated by rigid translations, rotations or tilts, and current adjustments without changing the manufactured coil shapes~\cite{wechsung_stochastic_2022}. 
Recent single-stage work simultaneously optimizes the plasma equilibrium boundary and stochastically optimizes the coils, and reports results comparable to stage-two stochastic optimization~\cite{gil_stochastic_2026}.

Most stochastic coil-optimization studies have focused on spatially varying coil-shape errors. Other perturbations have often been examined through sensitivity analysis or post-processing correction, including centroid displacement and tilt on CNT~\cite{hammond_experimental_2016}, rotations on W7-X~\cite{andreeva_analysis_2004,kislinger_correction_2005}, centroid displacement and tilt on CFQS~\cite{shimizu_consideration_2019,zhang_investigation_2025}, and rigid shifts, rotations, and current corrections in a posteriori optimization~\cite{wechsung_stochastic_2022}. These studies show that the relevant physics metric depends on the device. Examples include resonant fields and islands for NCSX and W7-X, flux-surface and neoclassical properties for CFQS, and magnetic-surface quality for LHD. The present work expands the set of perturbations treated directly during stage-two stochastic optimization.

\subsection{Contributions}
\label{sec:contributions}
This work provides an in-depth analysis of stage-two stochastic optimization and investigates perturbations to coil shapes (manufacturing errors), alignment and centroid position errors (assembly errors), and currents (errors from electromagnetic coupling). These errors could also come from deformation under the large electromagnetic loads. This work then explores combining this technique with recent improvements in stage-two coil optimization using the Augmented Lagrangian (AL) algorithm~\cite{gil_efficient_2026}. We demonstrate that the methods regularize the optimization landscape, converging to optima that are less sensitive to various error sources. We assess coil sets through the squared flux objective and its out-of-sample (OOS) distribution mimicking assembly and manufacturing errors in the coils, and rely on prior work relating squared flux fidelity to confinement quality~\cite{wechsung_single-stage_2022,wechsung_stochastic_2022} rather than performing comprehensive post-processing.

This study uses SIMSOPT~\cite{landreman_simsopt_2021} to model and optimize coils for several fixed plasma boundaries. Section~\ref{sec:Modeling} describes the filamentary representation~\cite{zhu_new_2017} and the applied perturbations. Section~\ref{sec:optimization model} describes the stochastic objective. Section~\ref{sec:results} presents results for the Landreman--Paul precise quasi-axisymmetric and quasi-helical cases~\cite{landreman_magnetic_2022} and for the NCSX c09r00 configuration, which was originally rescaled to target a nominal 0.5\,T on-axis magnetic field strength; hereafter the configurations are referred to as LPQA, LPQH, and NCSX respectively. The c09r00 equilibrium has also been used in permanent-magnet studies~\cite{zhu_perpendicular_2020,zhu_topology_2020,kaptanoglu_permanent_2022}.

\section{Modeling and Methodology}
\label{sec:Modeling}

\subsection{Coil Representation}
\label{sec:coil-representation}

Here we use the filamentary model of coils as used in previous work~\cite{zhu_new_2017}. The assumption is made that the coils have zero thickness and can be modeled as 1D curves carrying a current. A truncated Fourier series is used, where the $j$th coordinate of the $i$th coil is given by
\begin{equation}
    \Gamma^{i}_{j}(\theta) = a_{j,0}^{i} + \sum^{M}_{k=1} \left[a_{jk}^{i}\cos{(k\theta)} + b_{jk}^{i}\sin{(k\theta)}\right],
    \label{eq:coilgeo}
\end{equation}
where $\Gamma^{i}:[0,2\pi) \rightarrow \mathbb{R}^3$ and $j \in \{1,2,3\}$. The Fourier coefficients for the $i$th coil are stored in $x_i \in \mathbb{R}^{3(2M+1)}$. To preserve stellarator and field-period symmetries, only the base-coil coefficients are optimized, and the remaining coils are obtained by symmetry for a total of $2n_{\mathrm{fp}}N_b$ coils. The Fourier order is chosen here to be $M=16$ or $M=24$. The perturbations can be written generally as
\begin{equation}
    \tilde{\Gamma}^{i}_{j}(\theta) = \Gamma^{i}_{j}(\theta) + \epsilon,
\end{equation}
where $\epsilon$ represents manufacturing, centroid, orientation, or current error as defined below.
\subsection{Modeling Coil Position Perturbations}
\label{pert_model}

During manufacturing, the realized coil shape may differ from its design. We model this position error as a random periodic vector function $\epsilon_{\mathrm{p}}(\theta):[0,2\pi)\rightarrow\mathbb{R}^3$, whose three Cartesian components are independent Gaussian processes~\cite{rasmussen_gaussian_2006}. We distinguish the distribution used inside the stochastic objective from the distribution used for held-out sampling. The component standard deviations are $\sigt$ during training and $\sigs$ during OOS sampling. For $\alpha\in\{\mathrm{T},\mathrm{S}\}$, the stationary covariance of each Cartesian component is
\begin{equation}
    k_{\alpha}(d)=\sigma_{\alpha}^2
    \exp\left(-d^2/l_{\alpha}^2\right),
    \qquad d=\theta-\theta',
\end{equation}
where $l_{\alpha}$ is a correlation length in the dimensionless curve parameter $\theta$. Unless the curve is parameterized proportionally to arclength, a fixed $l_{\alpha}$ does not correspond to a fixed physical arclength along every coil. Periodicity is imposed by replacing $k_{\alpha}$ with
\begin{equation}
    \tilde{k}_{\alpha}(d)
    =\sum_{i\in\mathbb{Z}}k_{\alpha}(d+2\pi i),
\end{equation}
and truncating the sum after the exponentially small terms. 

At any fixed $\theta$, each Cartesian component is $\mathcal{N}(0,\sigma_{\alpha}^2)$. The pointwise magnitude $|\epsilon_\mathrm{p}|$ has a Maxwell distribution with
\begin{align}
    \mathbb{E}\!\left[\left|\epsilon_\mathrm{p}(\theta)\right|\right]
    &=2\sigma_{\alpha}\sqrt{\frac{2}{\pi}}
    \approx1.596\sigma_{\alpha},\\
    \sqrt{\mathbb{E}\!\left[\left|\epsilon_\mathrm{p}(\theta)\right|^2\right]}
    &=\sqrt{3}\,\sigma_{\alpha}
    \approx1.732\sigma_{\alpha}.
    \label{eq:pointwise-displacement}
\end{align}
Consequently, $\sigma_{\alpha}=5$\,mm corresponds to a mean pointwise displacement of $8.0$\,mm and an RMS displacement of $8.7$\,mm for the associated distribution. Spatial correlation changes how coherently neighboring points move but does not change these one-point values. To fix $l_{T}$ and $l_{S}$, we recreated the correlation-length scan of Wechsung \textit{et al.}~\cite{wechsung_stochastic_2022} and examined the resulting coils. Values $l<0.2$ produced highly oscillatory coils and an ill-conditioned covariance matrix leading to poor solutions, while $l\approx1$ left coil shapes qualitatively unperturbed. We therefore adopt the intermediate value $l_{T}=l_{S}=0.5$, consistent with prior work, which produces clear perturbations without excessive oscillations. This perturbation model is discussed in greater depth by Wechsung \textit{et al.}~\cite{wechsung_stochastic_2022}.

\subsection{Modeling Coil Centroid Perturbations}
\label{sec:centroid-perturbations}

During assembly of the coils, error may be introduced into the coil-centroid positions. For $\alpha\in\{\mathrm{T},\mathrm{S}\}$, the direction vector $\epsilon_\mathrm{dir}\in\mathbb{R}^3$ has independent $\mathcal{N}(0,1)$ components and is normalized to unit length. The signed amplitude is $\rho\sim\mathcal{N}(0,\sigma_{\alpha,\mathrm{c}}^2)$, where $\sigma_{\mathrm{T},\mathrm{c}}$ and $\sigma_{\mathrm{S},\mathrm{c}}$ are the training and OOS centroid scales. The centroid perturbation is the constant vector
\begin{equation}
    \epsilon_\mathrm{c}= \rho\,\frac{\epsilon_\mathrm{dir}}{\left|\epsilon_\mathrm{dir}\right|}.
\end{equation}

\subsection{Modeling Coil Orientation Perturbations}
\label{sec:orientation-perturbations}

Assembly error can also tilt a coil away from its intended orientation. We model this effect by rotating each coil about its own centroid, successively about the laboratory $x$, $y$, and $z$ axes. For $\alpha\in\{\mathrm{T},\mathrm{S}\}$, the three angles $\epsilon_\mathrm{a}=(\epsilon_1,\epsilon_2,\epsilon_3)$ are independent, with $\epsilon_j\sim\mathcal{N}(0,\sigma_{\alpha,j}^2)$. If $R(\epsilon_\mathrm{a})$ denotes the corresponding composite rotation and $c_i$ is the centroid of coil $i$, then
\begin{equation}
   \tilde{\Gamma}^{i}(\theta)=R(\epsilon_\mathrm{a})\left(\Gamma^i(\theta)-c_i\right)+c_i.
   \label{eq:orientation-perturbation}
\end{equation}
This centroid-based rigid-body convention separates translational and rotational assembly errors~\cite{andreeva_analysis_2004}. 


\subsection{Modeling Coil Current Perturbations}
\label{sec:current-perturbations}

During operation of a stellarator, the current can vary due to power-control error or induced currents from electromagnetic coupling, although the model assumes steady currents. For $\alpha\in\{\mathrm{T},\mathrm{S}\}$, the current errors are independent with $\epsilon_{I,i}\sim\mathcal{N}(0,\sigma_{\alpha,I}^2)$ for the training or OOS distribution. The current of the $i$th coil is then
\begin{equation}
    \tilde I_{i}=I_i+\epsilon_{I,i}.
\end{equation}

\section{Optimization Model}
\label{sec:optimization model}

The symmetric initial coil configuration, $\xzero$, is axisymmetric and scaled poloidally so as not to intersect the plasma surface. Perturbed initial configurations, $\xpert$, are also used and are discussed in Sec.~\ref{sec:pert_init}.

\subsection{Stochastic Objective}
\label{sec:stoch_obj}

The goal of stage-two stochastic optimization is to find coils that robustly reproduce the desired plasma while satisfying engineering constraints, even if they are significantly perturbed. We measure the accuracy of the coils matching the desired plasma boundary through the squared flux objective,
\begin{equation}
    \Jb(x,\epsilon) = \frac{1}{2}\int_S \left(B(x,\epsilon)\cdot\hat{n}\right)^2 \mathrm{d}S,
    \label{eq:squaredflux}
\end{equation}
evaluated in practice as a quadrature sum over discrete points on $S$, where $S$ is the fixed plasma equilibrium boundary. The plasma boundary for all geometries here is discretized at modest resolution over the full torus with $n_\phi = 64$ (toroidal) and $n_\theta = 16$ (poloidal) quadrature points. The stochastic objective used is risk-neutral and is formulated as
\begin{equation}
\label{eq:stochastic}
    \min_{x}\ \mathbb{E}\left[ \Jb(x,\epsilon)\right],
\end{equation}
where $x=(\{\Gamma_i\}_{i=1}^{N_b},\{I_i\}_{i=1}^{N_b})$ contains the Fourier coefficients and currents of the independent base coils. Stellarator and field-period symmetries map $x$ to the full set of $2n_{\mathrm{fp}}N_b$ coils before $B(x,\epsilon)$ is evaluated. The intended implementation holds one independent base-coil current fixed at $10^5$\,A, optimizes the other independent currents, and assigns symmetry-related currents from their corresponding base coils. Fixing one current excludes the zero-field solution, which would make the normal-flux objective vanish without generating the desired finite field.

Random perturbations are applied to calculate the perturbed magnetic field, $B(x,\epsilon)$, which is then used to calculate the flux. After $N$ perturbations, we apply the sample average approximation to obtain,
\begin{equation}
    \min_{x}\ \frac{1}{N}\sum_{i=1}^{N} \Jb(x,\epsilon_i).
    \label{eq:sampleavg}
\end{equation}

\subsection{Coil Penalty Objectives}
\label{sec:penalties}

In addition to the stochastic magnetic flux objective, we include several engineering penalty terms to ensure the coil solutions are physically realizable. They can also be viewed as regularization terms for the ill-posed problem~\eqref{eq:stochastic}. The total objective is a weighted sum of the magnetic flux objective and the following penalty terms, which we mention briefly but which are described in detail elsewhere, e.g.~\cite{kaptanoglu_reactor-scale_2025,zhu_new_2017}. Since coils are invariant under the stellarator and field-period symmetries relating each base coil to its symmetric images, it suffices to evaluate most coil-engineering-related penalties over only the base coils. The linking number and coil-coil distance penalties are the only terms that are evaluated on the full coil set, since e.g. coils at opposite ends can link when reproduced via symmetry.

\subsubsection{Coil Length}
\label{sec:penalty-length}
To prevent coils from becoming arbitrarily long, as well as to reduce cost and manufacturing difficulty, we penalize the total coil length above a threshold $L_{\max}$,
\begin{align}
    J_L &= \max\left(0,\, \sum_{i=1}^{N_b} L_i - L_{\max}\right)^2, \\ \notag L_i &= \int_0^{2\pi} \left|\frac{\mathrm{d}\Gamma_i}{\mathrm{d}\theta}\right| \mathrm{d}\theta.
\end{align}
The penalty acts on the total length summed over the base coils rather than on each coil individually.

\subsubsection{Coil-to-Coil Distance}
\label{sec:penalty-coil-coil}
This penalty helps avoid coil intersections and generally improves the practical feasibility of constructing the coils. We penalize coil pairs that come within a minimum distance $d_\mathrm{CC}$ of each other, while also preserving space for reactor equipment such as cooling systems and diagnostics.
\medmuskip=0mu
\thickmuskip=2mu
\begin{equation}
    J_\mathrm{CC} = \sum_{i=1}^{2n_{\mathrm{fp}}N_b} \sum_{j>i}^{2n_{\mathrm{fp}}N_b} \max(0,\, d_\mathrm{CC} - \min_{\theta, \theta'} \left|\Gamma_i(\theta) - \Gamma_j(\theta')\right|)^2.
\end{equation}
\medmuskip=4mu
\thickmuskip=4mu

\subsubsection{Coil-to-Surface Distance}
\label{sec:penalty-coil-surface}
To prevent coils from intersecting the plasma boundary, we penalize coils that come within a minimum distance $d_\mathrm{CS}$ of the surface $S$.
\begin{equation}
    J_\mathrm{CS} = \sum_{i=1}^{N_b} \max\left(0,\, d_\mathrm{CS} - \min_{\theta,\, x \in S} \left|\Gamma_i(\theta) - x\right|\right)^2.
\end{equation}

\subsubsection{Curvature}
\label{sec:penalty-curvature}
High-curvature regions of a coil are difficult to manufacture. We penalize curvature $\kappa_i(\theta)$ above a threshold $\kappa_{\max}$ in an $L^p$ sense with $p=2$, integrated with respect to arclength.
\begin{equation}
    J_{\kappa} = \sum_{i=1}^{N_b} \frac{1}{p}\int_0^{2\pi} \max\left(0,\, \kappa_i(\theta) - \kappa_{\max}\right)^{p} \left|\frac{\mathrm{d}\Gamma_i}{\mathrm{d}\theta}\right| \mathrm{d}\theta.
\end{equation}

\subsubsection{Mean Squared Curvature}
\label{sec:penalty-msc}
We also include a mean squared curvature penalty to globally discourage highly curved coil geometries.
\medmuskip=-1mu
\thickmuskip=-1mu
\begin{equation}
J_{\mathrm{MSC}} = \sum_{i=1}^{N_b} \max\left( 0,\, \frac{\int_0^{2\pi} \kappa_i(\theta)^2 \left|\frac{\mathrm{d}\Gamma_i}{\mathrm{d}\theta}\right| \mathrm{d}\theta}{\int_0^{2\pi} \left|\frac{\mathrm{d}\Gamma_i}{\mathrm{d}\theta}\right| \mathrm{d}\theta} - c_{\max} \right)^2.
\end{equation}
\medmuskip=4mu
\thickmuskip=4mu

\subsubsection{Arclength Variation}
\label{sec:penalty-arclength}
To ensure that each coil is discretized smoothly, an arclength variation penalty is incorporated. The arclength of the interval between each pair of adjacent discrete points is penalized against the mean interval arclength on that coil, keeping the arclength in each interval as constant as possible.
\begin{equation}
    J_A = \sum_{i=1}^{N_b} \frac{1}{L_i}\sum_{k=1}^{N_q} \left( \ell_{i,k} - \bar{\ell}_i \right)^2,
\end{equation}
where $\ell_{i,k}$ is the arclength of the $k$th interval on coil $i$, $\bar{\ell}_i$ is the mean interval arclength on that coil, $L_i$ is the coil length, and $N_q$ is the number of quadrature points.

\begin{table*}[t]
    \centering
    \caption{Numerical metrics used to evaluate nominal coil performance and
    robustness to coil perturbations. Since $\Jb$ and $\Joos$ have the same
    units, $Q$ is dimensionless.}
    \label{tab:vals}

    \setlength{\tabcolsep}{3pt}
    \begin{tabular}{@{}p{0.09\textwidth}p{0.31\textwidth}p{0.54\textwidth}@{}}
        \toprule
        Metric & Definition & Description \\
        \midrule
        $\Jb\ [\units]$
        &
        $\displaystyle
        \left\langle
        (B\cdot\hat{n})^2
        \right\rangle$
        &
        Plasma-surface-averaged squared flux (normal magnetic field) for the
        nominal optimized coil set.
        \\
        $\Joos\ [\units]$
        &
        $\displaystyle
        \frac{1}{N_{\mathrm{S}}}
        \sum_{i=1}^{N_{\mathrm{S}}}
        \left\langle
        (B_i\cdot\hat{n})^2
        \right\rangle$
        &
        Plasma-surface-averaged squared flux (mean normal field), averaged also over $N_{\mathrm{S}}$ perturbed
        coil sets representing manufacturing errors.
        \\
        $\Jnorm$
        &
        $\displaystyle
        \frac{
            \left\langle
            |B\cdot\hat{n}|
            \right\rangle
        }{
            \left\langle|B|\right\rangle
        }$
        &
        Mean absolute normal field on the plasma boundary, normalized by the
        mean field magnitude. This dimensionless metric is independent of the
        overall magnetic-field scale.
        \\
        $Q$
        &
        $\displaystyle
        \frac{\Jb}{\Joos}$
        &
        Dimensionless robustness ratio. Values near unity indicate little
        average degradation under coil perturbations, while smaller values
        indicate greater sensitivity.
        \\
        \bottomrule
    \end{tabular}
\end{table*}

\subsubsection{Linking Number}
\label{sec:penalty-linking}
The linking number $\mathcal{LK}$ counts the number of times two closed curves wind around each other. A non-zero linking number between two coils indicates that they are topologically linked, which is physically undesirable. We therefore include a penalty to discourage linked coil configurations.
\begin{align}
    J_{LK} &= \sum_{i=1}^{2n_{\mathrm{fp}}N_b}\sum_{j>i}^{2n_{\mathrm{fp}}N_b} \left|\mathcal{LK}\left(\Gamma_i, \Gamma_j\right)\right|,
\\ 
    \mathcal{LK}(\Gamma_i,\Gamma_j) &= \frac{1}{4\pi} \oint\!\!\oint \frac{\left(\mathrm{d}\Gamma_i \times \mathrm{d}\Gamma_j\right)\cdot\left(\Gamma_i - \Gamma_j\right)}{\left|\Gamma_i - \Gamma_j\right|^3},
\end{align}
so that $J_{LK} = 0$ for a fully unlinked coil set and $J_{LK} \geq 1$ whenever any pair is interlocked. The sum runs over the full coil set rather than the base coils alone. Since $\mathcal{LK}$ is integer-valued and piecewise constant, its analytic derivative is identically zero and the term contributes nothing to the descent direction; it is included in the objective value and therefore acts through the line search, where a trial step that links two coils raises $J_{\mathrm{coils}}$ and is rejected. This treatment mimics that in Giuliani's work~\cite{giuliani_direct_2024}. The penalty matters chiefly in the early stages of optimization, when coils may cross while navigating the optimization space; for all solutions reported here $J_{LK} = 0$ at convergence, so the term contributes neither to the objective value nor to the gradient at the reported minima.

\subsubsection{Cost Function}
\label{sec:cost-function}
The total objective is a weighted sum of each term, where $w_k$ are user-defined weights that balance the engineering constraints against the magnetic flux objective.
\begin{align}
    J_{\mathrm{coils}} = \mathbb{E}\left[\Jb\right] &+ w_L J_L + w_{CC} J_{CC} + w_{CS} J_{CS} 
    \\ \notag &+ w_{\kappa} J_{\kappa} + w_{MSC} J_{MSC} \\ \notag
    &+ w_A J_{A} + w_{LK} J_{LK}.
\end{align}
The thresholds and weights for each coil penalty objective are defined in Appendix~\ref{sec:appendix} for each stellarator studied, along with other geometric parameters. All objectives share the design vector $x$, which contains the Fourier coefficients and currents of the base coils.

\subsection{Stochastic AL Optimization}
\label{sec:stochastic-al}

The AL numerical method works by solving the optimization repeatedly while automatically tuning the weights at each step in the sequence to ensure each objective stays within a defined constraint~\cite{gil_efficient_2026,conlin_stellarator_2024,nocedal_numerical_2006,conn_trust_2000}. The desired result is a set of coils that satisfies every constraint, which is challenging to impose when nonconvex constraints are present. The numerical method is described in detail in Gil \textit{et al.}~\cite{gil_efficient_2026}.

We implement the stochastic objective, equation~\eqref{eq:stochastic}, to combine the AL and stochastic optimization methods. This is done to gain the benefits of a more advanced numerical solver without having to tune weights manually, and to investigate the ability of the stochastic objective to regularize the optimization landscape and yield more robust solutions. This method was applied to the most coil-sensitive stellarator investigated in this work, as manual weight and threshold tuning is particularly challenging for that configuration.

\subsection{Evaluation Criteria}
\label{sec:eval}

The optimizer's performance is evaluated using the metrics defined in Table~\ref{tab:vals}, which give a general overview of optimization performance and engineering constraint behavior. In particular, we report our coil robustness metric $Q$, which compares the ratio of $J_B$ evaluated at the final solution from optimization (the ``nominal'' solution), and $J_S$, the squared flux averaged over $N_S$ perturbation realizations applied to that nominal solution. If $Q = 1$, $J_S = J_B$, i.e. there is no degradation of the squared flux under coil perturbations. If $Q \to 0$, $J_S \gg J_B$, meaning the coils are very sensitive to perturbations.

\subsubsection{Out-of-Sample Evaluation}
\label{sec:oos}

The primary quantitative measure is the OOS evaluation. After optimization, each stochastic or deterministic coil set is subjected to held-out perturbations with component standard deviation $\sigs$ and correlation length $l_{\mathrm{S}}$. The squared flux is evaluated for $N_{\mathrm{S}}=1000$ independently sampled perturbed coil sets. Their distribution quantifies how the optimized design responds to the specified manufacturing or assembly error model.

Unless otherwise specified in specific decoupled variable scans, we will assume $\sigs=\sigt$ and $l_{\mathrm{S}}=l_{\mathrm{T}}$. The stochastic solution is therefore evaluated against the same distribution scale used during training. A deterministic run has no $\sigt$ because it is trained without perturbations, but it is evaluated at the same $\sigs$ as the stochastic solution to which it is compared. In Sec.~\ref{sec:sigma-mesh-discussion}, we also report decoupled OOS scans, holding the coil set and $\sigt$ fixed while varying $\sigs$. A fixed random seed provides identical OOS realizations across run modes.

We evaluate the OOS flux distributions using traditional symmetric initial coil-set configurations, $\xzero$, and randomly perturbed initial configurations, $\xpert$, explained further in Sec.~\ref{sec:pert_init}. These distributional evaluation methods have been demonstrated in prior work~\cite{liu_influence_2026,wechsung_stochastic_2022}.

\subsubsection{Perturbed Initial Guess Evaluation}
\label{sec:pert_init}

To investigate the optimization landscape of each stellarator configuration, we use an ensemble of $N_0$ perturbed initial coil sets $\xpert$ that use the same distribution as the perturbations applied during training, $\sigma_T$ and $l_T$. These same initial configurations are supplied to the deterministic and stochastic optimizers. An OOS evaluation is then run from each optimized solution, and the resulting distributions are compared. This analysis tests whether the robustness improvement persists across different basins of attraction.

\section{Results and Discussion}
\label{sec:results}

\begin{figure}
    \centering
    \includegraphics[width=0.45\textwidth]{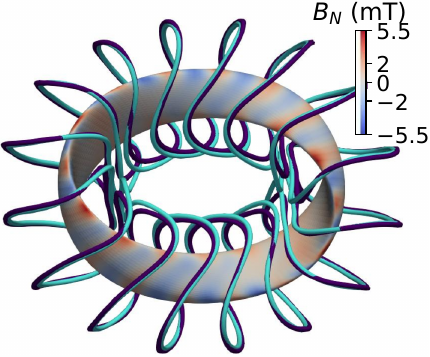}
    \par\vspace{3pt}
    \includegraphics[width=0.45\textwidth]{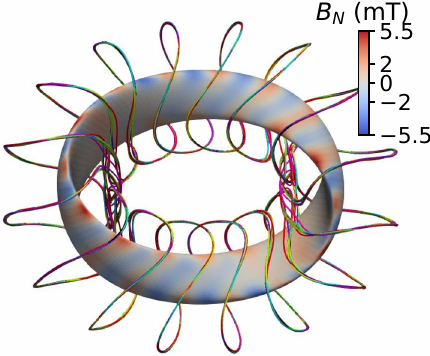}
    \par\vspace{3pt}
    \includegraphics[width=0.45\textwidth]{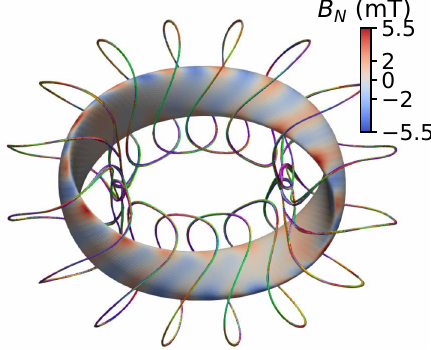}
    \caption{LPQA coil geometry. The top plot shows deterministic (dark purple) and 50S (cyan) coil solutions overlaid on the plasma boundary colored by the normal field component $B_n$. The middle plot shows converged coil sets from the ensemble of perturbed initial configurations $\xpert$ under deterministic optimization, with each color denoting a different initialization. The bottom plot shows the same ensemble under 50S optimization at the same scale.}
    \label{fig:qa_geom}
\end{figure}

\subsection{Optimizer Details}
\label{sec:optimizer-details}

Deterministic and stochastic objectives are minimized with L-BFGS-B~\cite{virtanen_scipy_2020}, using the analytic gradient, 300 correction pairs, and a convergence tolerance of $10^{-15}$ so that runs terminate on the iteration budget rather than early stopping. Three run modes are compared. These are deterministic optimization of the unperturbed objective and stochastic optimization with $N = 4$ and $N = 50$ perturbation samples, denoted 4S and 50S. All three use identical penalty weights, thresholds, and perturbation parameters. Numerical results are reported for the deterministic and $N = 50$ cases only. The $N = 4$ case appears in the figures to illustrate sample-count dependence. The value $N = 50$ was selected from a confidence-interval estimate on the sample-average objective, placing it at the low end of acceptable variance. Sampling at larger $N$ produced negligible changes. The larger iteration budget for the stochastic runs (2000 instead of 1000) accommodates the slower convergence of the sample-averaged objective. Gradient norms at termination were verified for every run for both methods, indicating that the difference does not account for the reported performance gap.

For stochastic training, $\sigt=1$\,cm is used, and halved to $\sigt=5$\,mm if this perturbation size produced only poor final solutions.

\begin{figure*}
    \centering
    \includegraphics[width=0.47\textwidth]{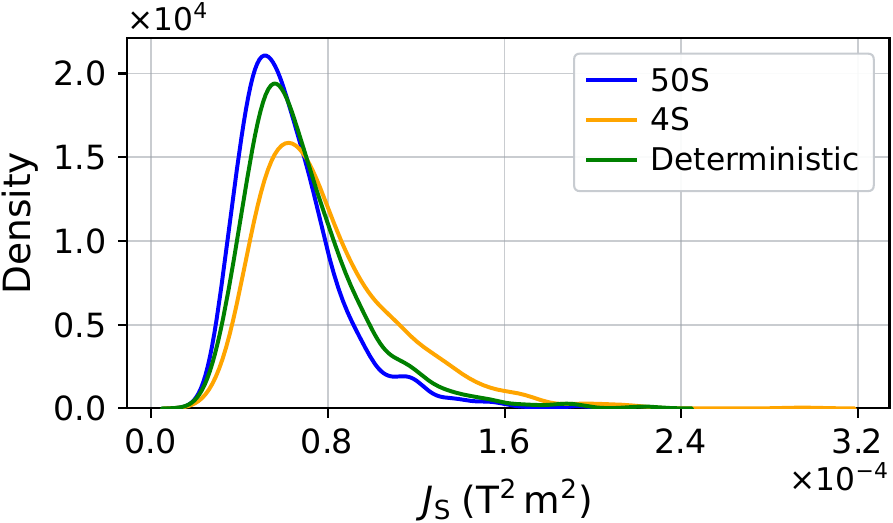}\hfill
    \includegraphics[width=0.47\textwidth]{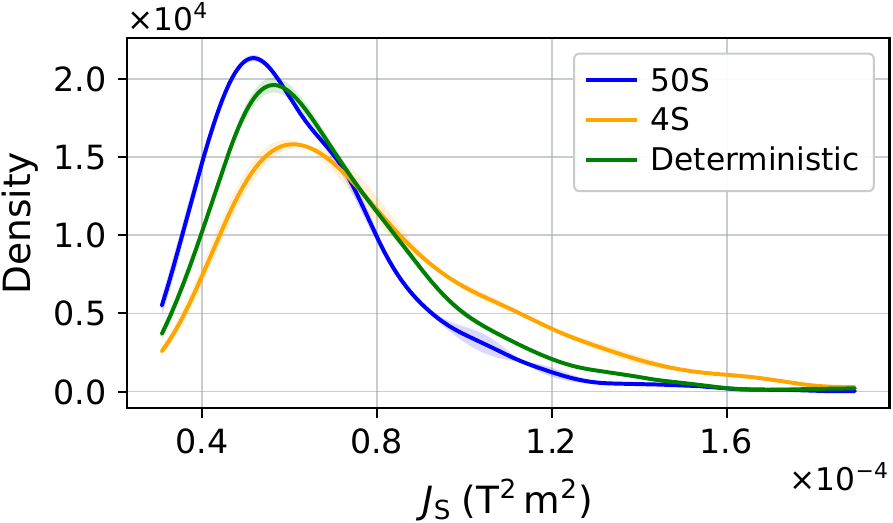}
    \par\vspace{6pt}
    \includegraphics[width=0.47\textwidth]{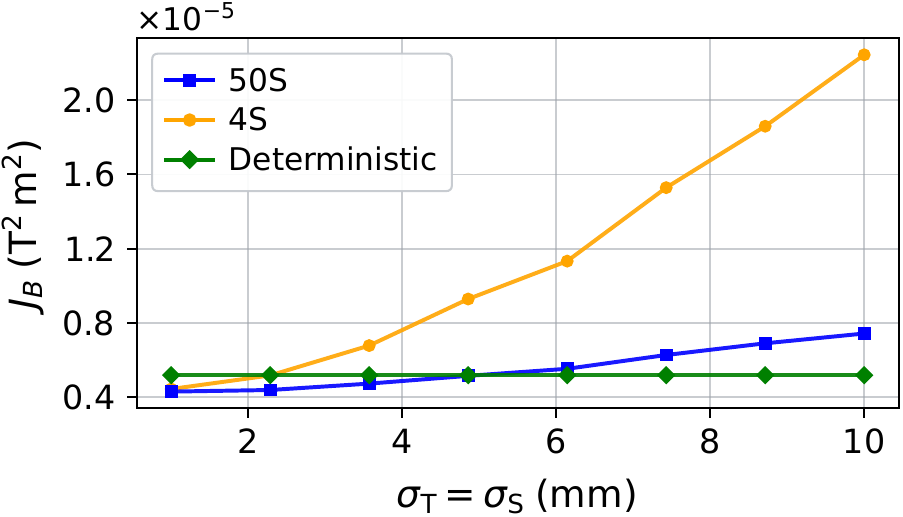}\hfill
    \includegraphics[width=0.47\textwidth]{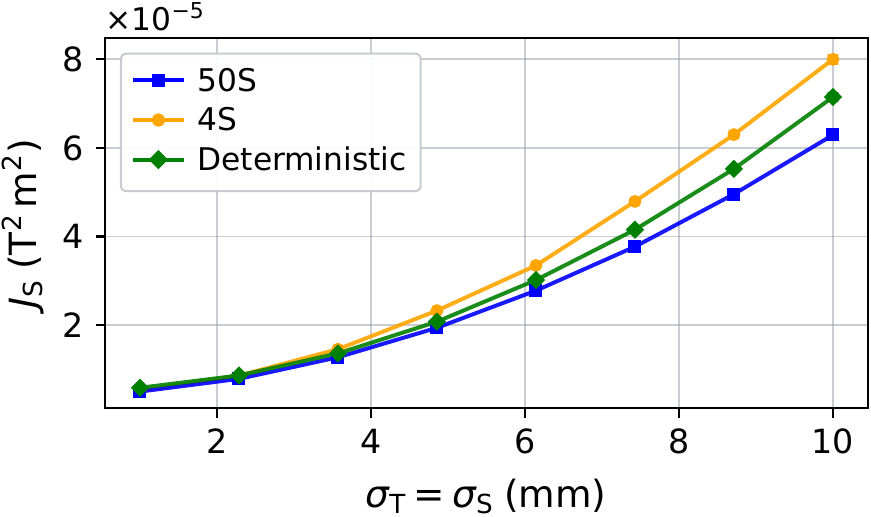}
    \caption{LPQA out-of-sample statistics comparing the deterministic, 4S and 50S solutions. The top-left plot shows squared-flux OOS distributions from the symmetric initialization $\xzero$. The top-right plot shows the same distributions averaged over the ensemble of perturbed initializations $\xpert$, with shaded bands giving the standard deviation across initializations. The bottom row shows the nominal objective $\Jb$ and mean OOS flux $\Joos$ over the matched scan $1\,\mathrm{mm}\leq\sigt=\sigs\leq1\,\mathrm{cm}$.}
    \label{fig:qa_stats}
\end{figure*}

\subsection{Results for Coil Manufacturing Perturbations}
\label{sec:pos_results}

\subsubsection{LPQA}
\label{sec:lpqa-results}

\begin{table}[tb]
    \centering
    \caption{Numerical metrics for the deterministic (Det) and 50S LPQA optimizations with $\sigt=\sigs=1$\,cm. The 50S optimization reduces the mean perturbed-field
    error and increases the robustness ratio $Q$, at the cost of a modest
    increase in the nominal-field error.}
    \label{tab:stdnumvals_lpqa}

    \setlength{\tabcolsep}{12pt}
    \begin{tabular}{@{}lcc@{}}
        \toprule
        Metric & Det & 50S \\
        \midrule
        $\Jb\;(\units)$    & $\mathbf{5.7\times10^{-6}}$ & $7.7\times10^{-6}$ \\
        $\Joos\;(\units)$  & $6.9\times10^{-5}$ & $\mathbf{6.3\times10^{-5}}$ \\
        $\Jnorm$           & $\mathbf{3.3\times10^{-3}}$ & $3.6\times10^{-3}$ \\
        $Q$                & $0.0812$             & $\mathbf{0.122}$              \\
        \bottomrule
    \end{tabular}
\end{table}

The Landreman--Paul precise quasi-axisymmetric (LPQA) stellarator configuration~\cite{landreman_magnetic_2022} serves as our first test case. LPQA configurations are known to possess a favorable optimization landscape for coil design, as the axisymmetric-like structure of the magnetic field tends to produce smoother, less geometrically complex coil solutions. This makes LPQA a useful baseline for assessing whether the stochastic optimization framework confers additional robustness benefits even where the deterministic solution is already relatively well conditioned.

As shown in the top panel of Figure~\ref{fig:qa_geom}, the deterministic and stochastic coil solutions are geometrically distinct, occupying two separate local minima in the optimization space, with the corresponding numerical values given in Table~\ref{tab:stdnumvals_lpqa}. The two solutions nevertheless achieve comparable squared flux fidelity, with $\Jb$ differing by roughly 36\% --- a small separation relative to the order-of-magnitude variation seen across configurations, and consistent with the better-conditioned LPQA landscape in which both optimizers reach nearby minima. Under OOS sampling, the stochastic solution consistently demonstrates superior robustness. The OOS flux $\Joos$ is reduced relative to the deterministic baseline, and the quality number $Q$ improves from $0.0812$ to $0.122$, an improvement of approximately $50.2\%$.

The OOS analysis in the top left panel of Figure~\ref{fig:qa_stats} illustrates this behavior. The 50S solution produces a distribution of perturbed squared-flux values concentrated below the deterministic baseline. For LPQA the separation across perturbed initializations is complete. Across $N_0 = 16$ perturbed initializations, every stochastic run yields a lower mean out-of-sample flux than every deterministic run, with ranges $[6.3,6.4]\times10^{-5}$ and $[6.8,7.1]\times10^{-5}\ \units$, respectively. This corresponds to an 8.5\% reduction in ensemble mean and is suggestive of a regularization effect on the optimization landscape.

The converged coil sets from each initialization are shown in the middle and bottom panels of Figure~\ref{fig:qa_geom}. The bottom row of Figure~\ref{fig:qa_stats} shows that the 50S solution maintains its advantage across the range $1\,\mathrm{mm}\leq\sigt=\sigs\leq1\,\mathrm{cm}$. The 4S runs do not share this behavior and are discussed in Sec.~\ref{sec:samplecount}.

\subsubsection{NCSX}
\label{sec:ncsx-results}
The National Compact Stellarator Experiment (NCSX) configuration provides a physically motivated and historically significant test case for the stochastic coil design framework. NCSX was a compact quasi-axisymmetric stellarator designed at the Princeton Plasma Physics Laboratory whose construction was canceled in 2008, in large part due to cost overruns driven by the tight manufacturing tolerances required by its complex coil geometry~\cite{neilson_lessons_2010}.

\begin{table}[tb]
    \centering
    \caption{Numerical metrics for the deterministic (Det) and 50S NCSX optimizations with $\sigt=\sigs=1$\,cm. The 50S optimization modestly reduces the mean
    perturbed-field error and increases the robustness ratio $Q$, while
    leaving the nominal metrics nearly unchanged.}
    \label{tab:stdnumvals_ncsx}

    \setlength{\tabcolsep}{12pt}
    \begin{tabular}{@{}lcc@{}}
        \toprule
        Metric & Det & 50S \\
        \midrule
        $\Jb\;(\units)$   & $\mathbf{2.6\times10^{-4}}$ & $\mathbf{2.6\times10^{-4}}$ \\
        $\Joos\;(\units)$ & $4.9\times10^{-4}$ & $\mathbf{4.8\times10^{-4}}$ \\
        $\Jnorm$          & $1.7\times10^{-2}$ & $\mathbf{1.6\times10^{-2}}$ \\
        $Q$               & $0.512$              & $\mathbf{0.535}$              \\
        \bottomrule
    \end{tabular}
\end{table}

\begin{figure}
    \centering
    \includegraphics[width=0.39\textwidth]{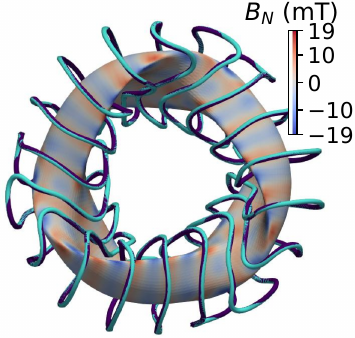}
    \par\vspace{1pt}
    \includegraphics[width=0.39\textwidth]{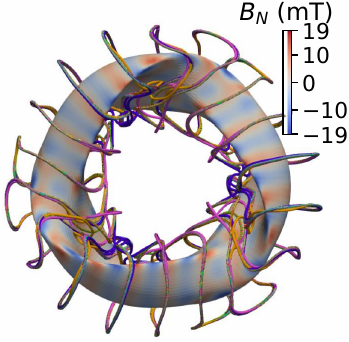}
    \par\vspace{1pt}
    \includegraphics[width=0.39\textwidth]{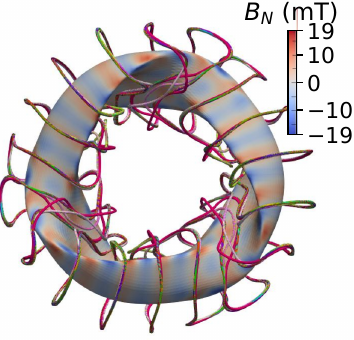}
    \caption{NCSX coil geometry. The top plot overlays the deterministic (dark purple) and 50S (cyan) coil solutions on the plasma boundary, colored by the normal field component $B_n$. The middle and bottom plots show converged coil sets from the ensemble of perturbed initial configurations $\xpert$ under deterministic and 50S optimization, respectively. Each color denotes a different initialization.}
    \label{fig:ncsx_geom}
\end{figure}

\begin{figure*}[tb]
    \centering
    \includegraphics[width=0.47\textwidth]{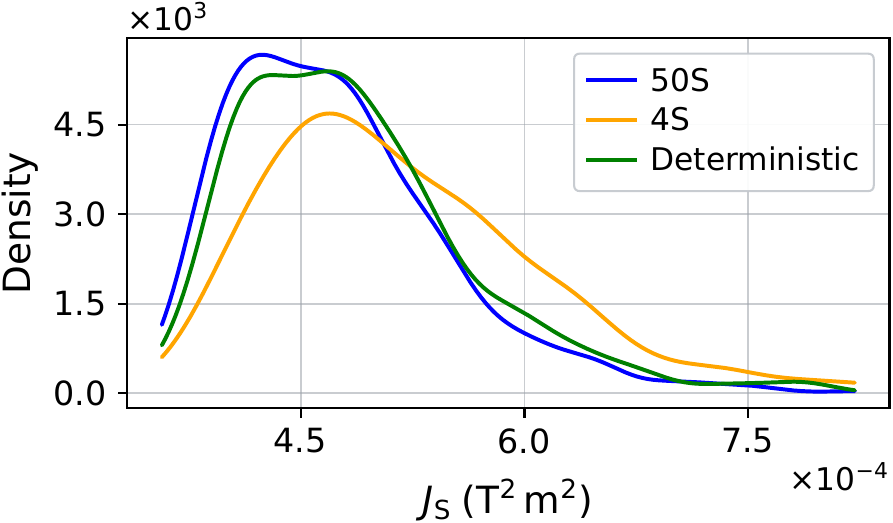}\hfill
    \includegraphics[width=0.47\textwidth]{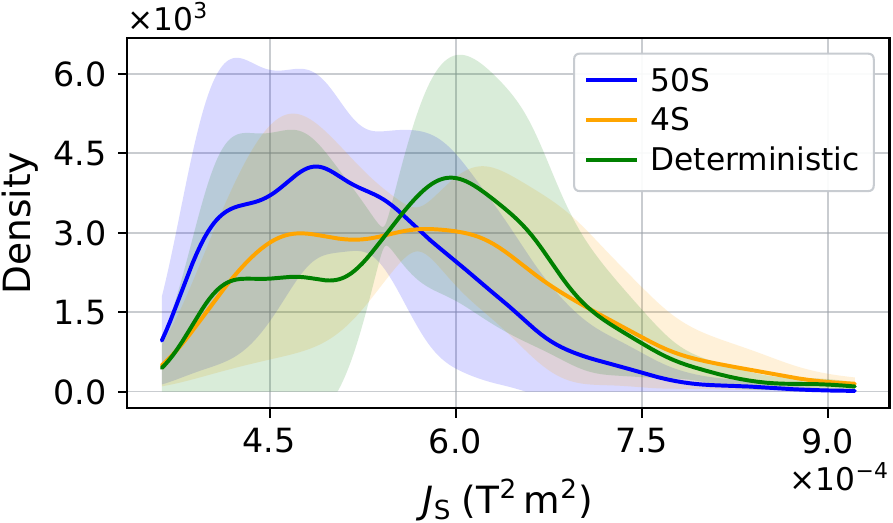}
    \par\vspace{6pt}
    \includegraphics[width=0.47\textwidth]{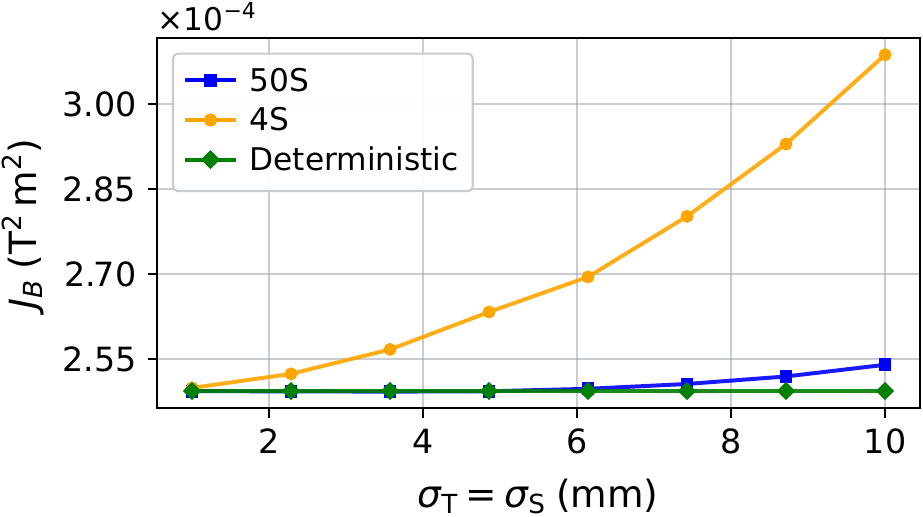}\hfill
    \includegraphics[width=0.47\textwidth]{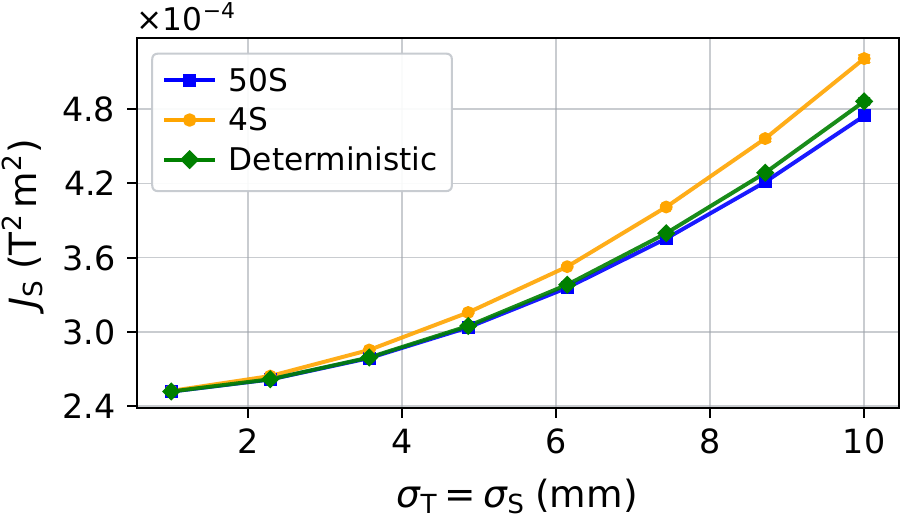}
    \caption{NCSX out-of-sample statistics comparing the deterministic, 4S and 50S solutions. The top-left plot shows squared-flux OOS distributions from the symmetric initialization $\xzero$. The top-right plot shows the corresponding distributions averaged over the ensemble of perturbed initializations $\xpert$, with shaded bands giving the standard deviation across initializations. The bottom row shows $\Jb$ and $\Joos$ over the matched scan $1\,\mathrm{mm}\leq\sigt=\sigs\leq1\,\mathrm{cm}$.}
    \label{fig:ncsx_stats}
\end{figure*}

NCSX adopted a 1.5\,mm modular-coil tolerance after error-field calculations showed that larger deviations could generate resonant islands occupying an unacceptable fraction of the usable toroidal flux~\cite{brooks_coil_2003}. This makes NCSX a valuable candidate for stochastic coil optimization because sensitivity of its magnetic surfaces to geometric error was evaluated directly at the project level. 

The deterministic and stochastic coil solutions are compared in the top plot of Figure~\ref{fig:ncsx_geom}, and the corresponding OOS distributions from the symmetric initialization are shown in the top-left plot of Figure~\ref{fig:ncsx_stats}. Table~\ref{tab:stdnumvals_ncsx} shows that the nominal metrics are nearly unchanged. The 50S solution lowers $\Joos$ by only 2.3\%, while $Q$ increases from 0.512 to 0.535, an improvement of approximately 4.5\%. Thus the principal NCSX result is that stochastic training provides very little benefit for the symmetric starting configuration.

\begin{figure*}
    \centering
    \includegraphics[width=0.33\textwidth]{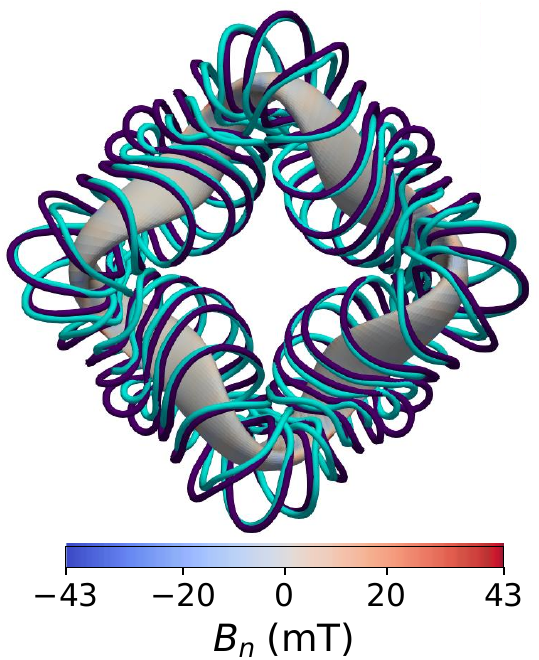}\hfill
    \includegraphics[width=0.33\textwidth]{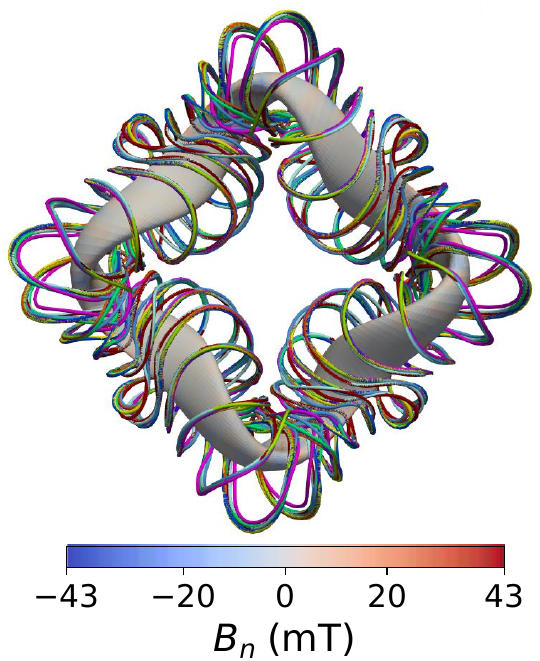}
    \hfill
    \includegraphics[width=0.33\textwidth]{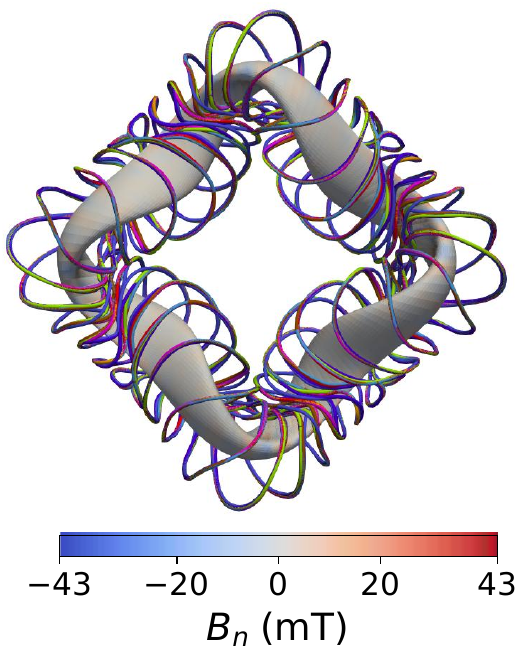}
    \caption{LPQH coil geometry. The left plot overlays deterministic (dark purple) and 50S (cyan) coil solutions on the plasma boundary, colored by the normal field component $B_n$. The middle and right plots show converged coil sets from the ensemble of perturbed initial configurations $\xpert$ under deterministic and 50S optimization, respectively, at the same scale. The deterministic runs exhibit much more variation in the final coil sets.}
    \label{fig:qh_geom}
\end{figure*}

\begin{figure*}
    \centering
    \includegraphics[width=0.45\textwidth]{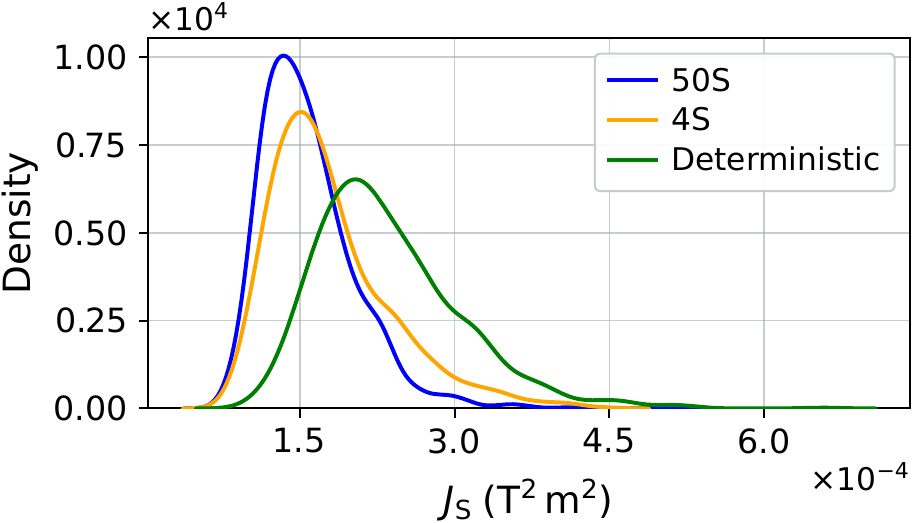}\hfill
    \includegraphics[width=0.45\textwidth]{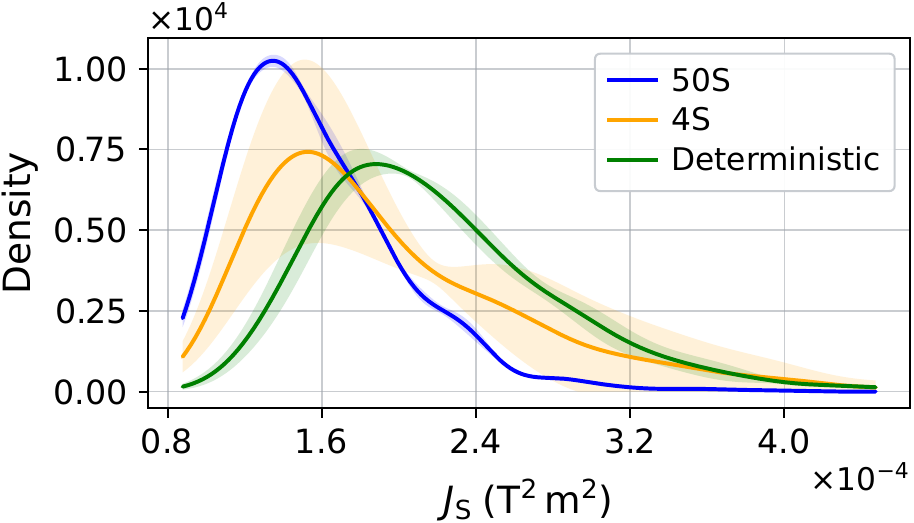}
    \par\vspace{6pt}
    \includegraphics[width=0.45\textwidth]{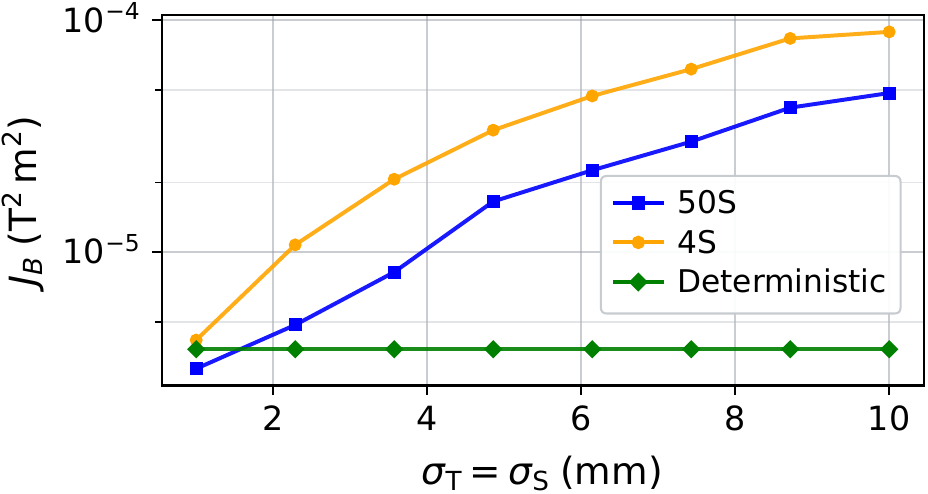}\hfill
    \includegraphics[width=0.45\textwidth]{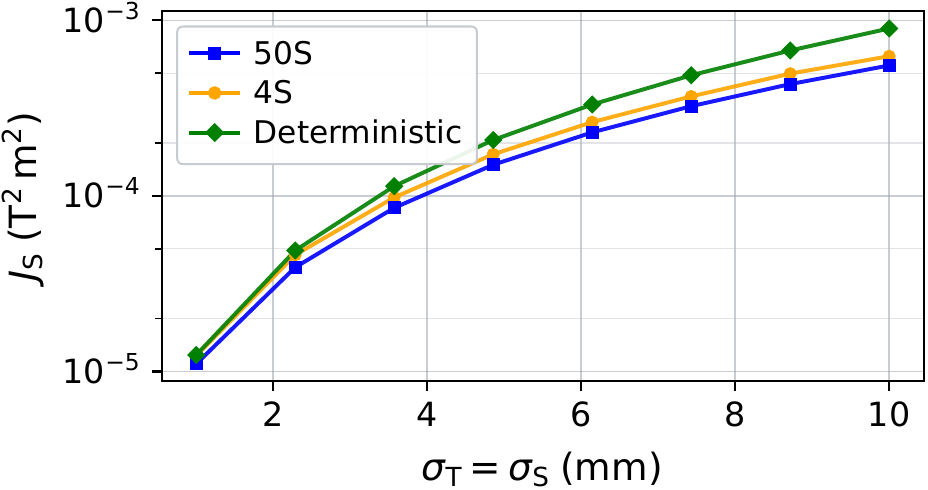}
    \caption{LPQH out-of-sample statistics comparing the deterministic, 4S and 50S solutions. The top-left plot shows squared-flux OOS distributions from the symmetric initialization $\xzero$. The top-right plot shows the corresponding distributions averaged over the ensemble of perturbed initializations $\xpert$, with shaded bands giving the standard deviation across initializations. The bottom row shows $\Jb$ and $\Joos$ over the matched scan $1\,\mathrm{mm}\leq\sigt=\sigs\leq1\,\mathrm{cm}$.}
    \label{fig:qh_stats}
\end{figure*}

Across the ensemble of $N_0 = 10$ perturbed initial guesses, there are significant differences in the final coil solutions on the inboard side.
The visually similar final coil sets in Figure~\ref{fig:ncsx_geom} are the single symmetric-start comparison, whereas the broad top-right bands in Figure~\ref{fig:ncsx_stats} aggregate many separately optimized perturbed initial guesses. These results show that both objectives can reach multiple nearby basins.

The scans in the bottom row of Figure~\ref{fig:ncsx_stats} show that the 50S $\Joos$ remains at or below the deterministic $\Joos$, with the two nearly coincident at the smallest values of $\sigt=\sigs$. Taken together, the NCSX results support only minimal robustness improvement. Our interpretation is that there are many local minima for this NCSX coil design, but that each of the local minima exhibits very similar coil metrics and very similar squared flux metric; these local minima seem to correspond to slightly different combinations of high curvature and near-overlap of the coils in the inboard side of the plasma.

\subsubsection{LPQH}
\label{sec:lpqh-results}

\begin{table}[tb]
    \centering
    \caption{Numerical metrics for the deterministic (Det) and 50S LPQH optimizations. The 50S optimization reduces the mean perturbed-field
    error and substantially increases the robustness ratio $Q$, at the cost
    of a larger nominal-field error.}
    \label{tab:stdnumvals_qh}

    \setlength{\tabcolsep}{12pt}
    \begin{tabular}{@{}lcc@{}}
        \toprule
        Metric & Det & 50S \\
        \midrule
        $\Jb\;(\units)$   & $\mathbf{3.3\times10^{-6}}$ & $1.8\times10^{-5}$ \\
        $\Joos\;(\units)$ & $2.4\times10^{-4}$ & $\mathbf{1.6\times10^{-4}}$ \\
        $\Jnorm$          & $\mathbf{1.2\times10^{-3}}$ & $2.9\times10^{-3}$ \\
        $Q$               & $0.0140$             & $\mathbf{0.116}$              \\
        \bottomrule
    \end{tabular}
\end{table}

The Landreman--Paul precise quasi-helically symmetric (LPQH) stellarator configuration~\cite{landreman_magnetic_2022} presents the most challenging coil optimization problem of the three configurations studied here, as the squared flux sensitivity appears orders of magnitude higher than the other designs; we demonstrate below that it benefits most from a stochastic approach. 

The helical symmetry of the LPQH configuration imposes stringent requirements on the coil geometry, and prior work has established that filamentary LPQH coil optimization is highly sensitive to the choice of penalty weights and thresholds in the cost function~\cite{wiedman_coil_2024}. Accordingly, the matched comparison uses $\sigt=\sigs=5$\,mm instead of the 1\,cm used for LPQA and NCSX.

The deterministic solution is compared with Wiedman \textit{et al.}~\cite{wiedman_coil_2024} for an order-of-magnitude agreement rather than a direct comparison, as we used a slightly different boundary file and the primary goal here is to compare our own equivalent deterministic and stochastic optimizations. A definitive numerical comparison therefore requires rerunning with the archived Wiedman boundary. Figure~\ref{fig:qh_geom} shows that the stochastic and deterministic optimizations locate geometrically distinct minima. Table~\ref{tab:stdnumvals_qh} gives the numerical comparison. The 50S solution has a larger nominal $\Jb$ but a smaller absolute $\Joos$, and $Q$ improves from 0.0140 to 0.116, a factor of 8.3.

The OOS distributions in the top row of Figure~\ref{fig:qh_stats} reflect this clearly. The stochastic distribution is concentrated at significantly lower squared-flux values under perturbation, with a narrower distribution indicating reduced variance across manufacturing-error realizations. As with LPQA, the separation across perturbed initializations is complete. The stochastic ensemble spans $[1.56,1.59]\times10^{-4}$ against a deterministic range of $[1.99,2.31]\times10^{-4}\ \units$, a 28.7\% reduction in ensemble mean. The coefficient of variation across initializations falls from 4.4\% to 0.75\%, showing that stochastic optimization here exhibits a strong regularization effect on the landscape.
The scan results in the bottom row of Figure~\ref{fig:qh_stats} confirm that the 50S solution consistently maintains lower $\Joos$ than the deterministic baseline, including at perturbation magnitudes above the 5\,mm training scale, indicating that the robustness benefit extends beyond the perturbation size used in training. Among the converged solutions shown in the middle and right panels of Figure~\ref{fig:qh_geom}, the stochastic runs on the right cluster more tightly than the deterministic ones in the middle, consistent with the ensemble statistics reported above.

\begin{figure}
\centering
    \includegraphics[width=.45\textwidth]{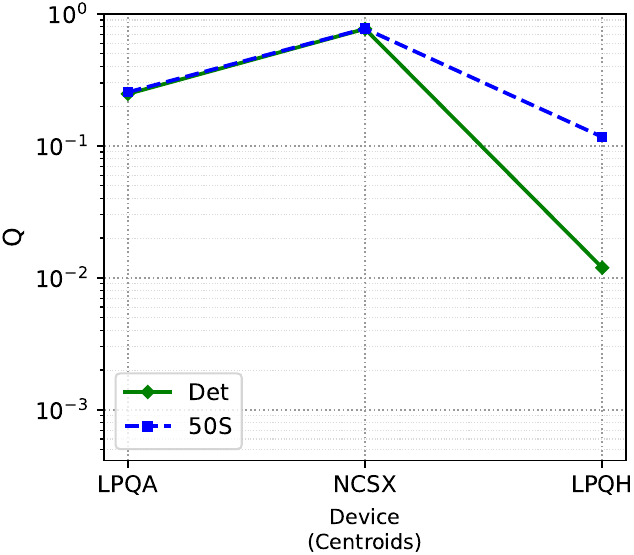}
    \par\vspace{1pt}
    \includegraphics[width=.45\textwidth]{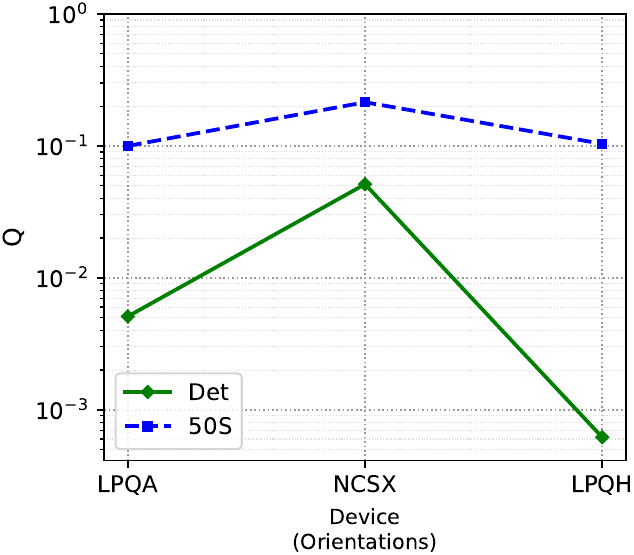}
    \par\vspace{1pt}
    \includegraphics[width=.45\textwidth]{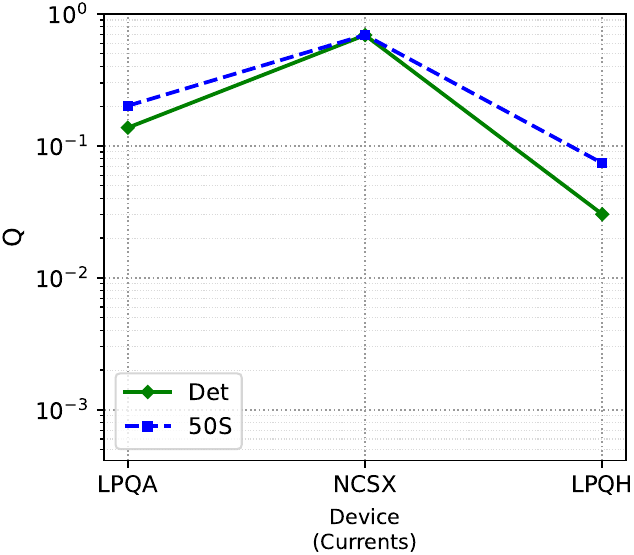}
    \caption{Quality number $Q$ for the deterministic and 50S solutions across all three stellarator configurations and the three additional perturbation models. The top, middle, and bottom plots show centroid, orientation, and current perturbations, respectively.}
    \label{fig:addit_pert_Q}
\end{figure}

\subsection{Rigid-Body and Current Perturbations}
\label{sec:rigidbody}

Having covered the coil position perturbations, we now turn to results for rigid-body and current perturbations, given the importance of these on the magnetic field discussed in Sec.~\ref{sec:Introduction}. As with the coil position perturbations, deterministic and 50S optimizations are run for the LPQA, LPQH, and NCSX stellarators, optimizing for centroid, orientation, and current perturbations separately.

 \begin{figure}
    \centering
    \includegraphics[width=0.47\textwidth]{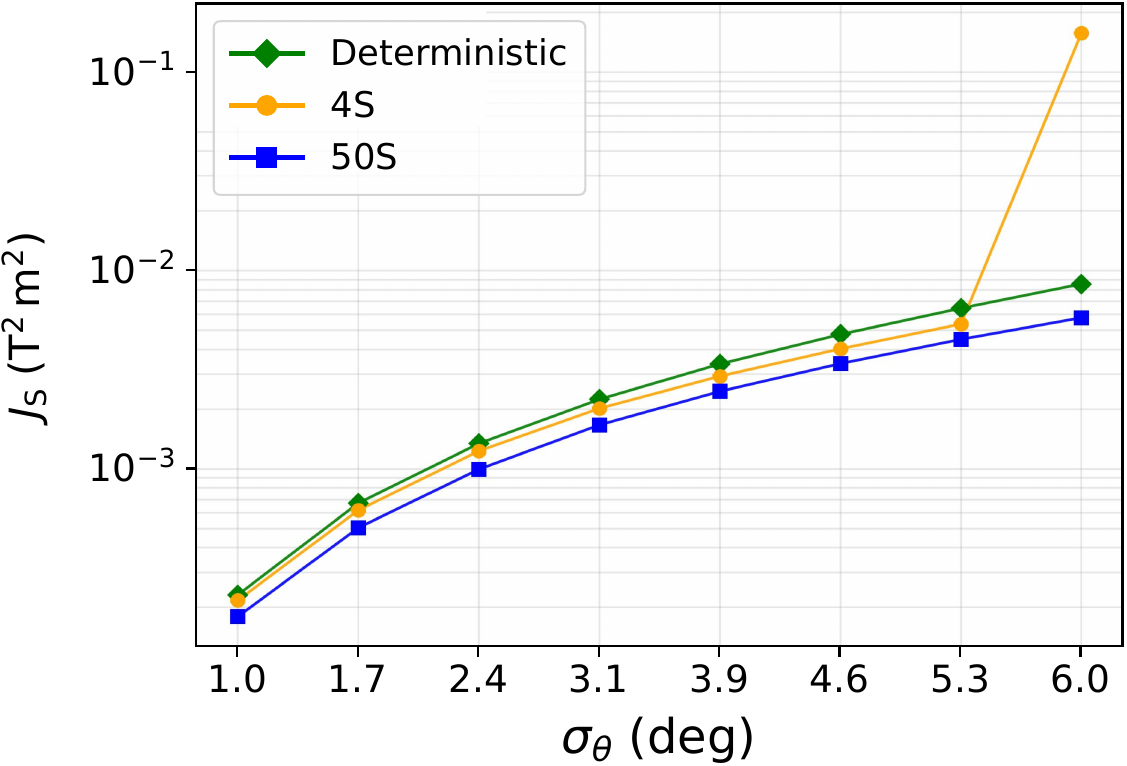}
    \par\vspace{1pt}
    \includegraphics[width=0.47\textwidth]{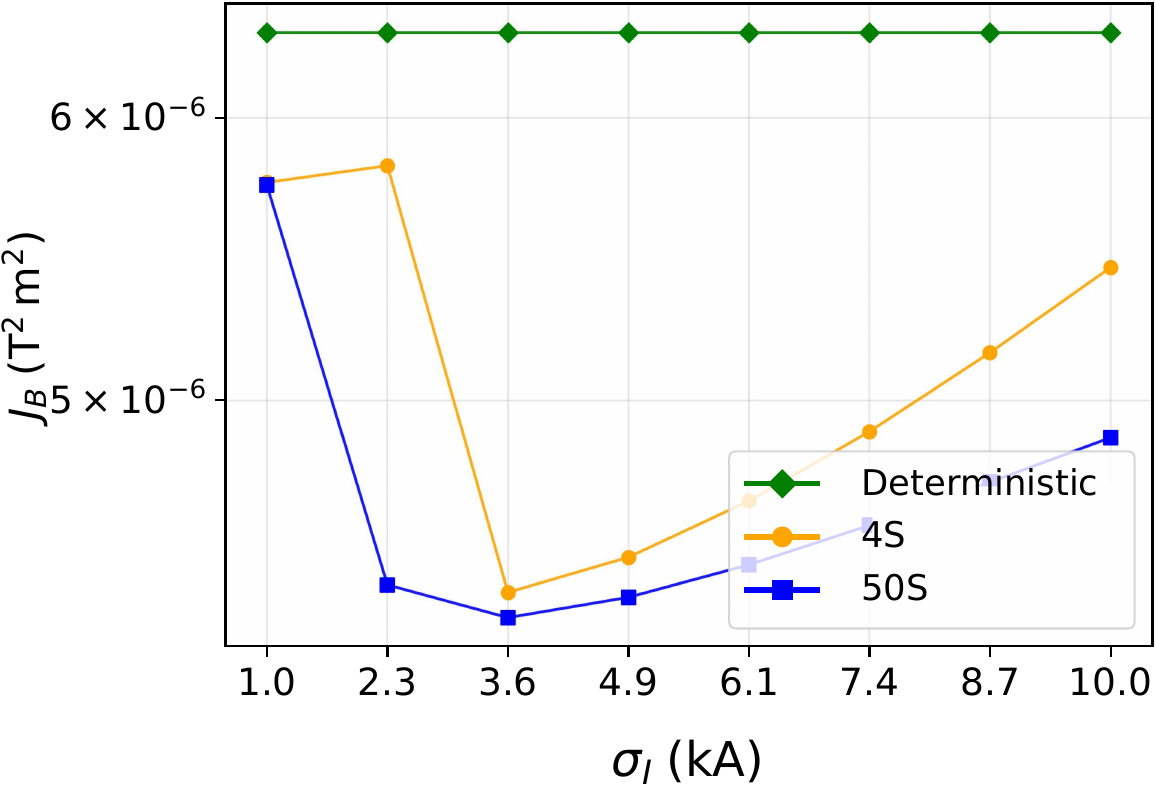}
    \caption{The top plot shows mean out-of-sample squared flux $\Joos$ for stochastic orientation optimization of LPQH as a function of orientation perturbation size. The bottom plot shows nominal squared flux $\Jb$ for stochastic current optimization of LPQA as a function of current perturbation size.}
    \label{fig:jf_orient_current}
\end{figure}

The stochastic training scales are 1\,cm for the centroid perturbation, $5^\circ$ about each axis for the orientation perturbation, and 10\,kA for the current perturbation. These scales are distribution parameters used to stress-test the optimizer and are not equivalent across perturbation types.

Using these perturbation sizes, the $Q$ values for the deterministic and stochastic optimizations under centroid, orientation, and current perturbations are shown in Figure~\ref{fig:addit_pert_Q}. Analysis of the other evaluation criteria defined in Sec.~\ref{sec:eval} yields results consistent with the findings for the coil position perturbations, and they are omitted here to avoid redundancy, but all 35 additional plots are available in the accompanying data repository (see Sec.~\ref{sec:data_availability}).

First, the deterministic coils are already robust to the centroid perturbations, so the stochastic optimization has little room to improve for the LPQA and NCSX stellarators; the differences in $Q$ between the deterministic and stochastic results are correspondingly small. This is consistent with the finding in~\cite{zhang_investigation_2025}, which found negligible changes to the effective helical ripple under centroid displacements at or below 10\,mm. However, consistent with the finding that LPQH is highly sensitive, this configuration's $Q$ value improves by an order of magnitude.

The orientation perturbation size is large enough that the stochastic optimizer is close to producing only poor solutions. However, from the deterministic to the stochastic optimizer the $Q$ value improves significantly, most pronounced in the LPQH case where it improves by nearly three orders of magnitude. Combined with satisfactory $\Jb$ values for the stochastic optimization, shown in the top panel of Figure~\ref{fig:jf_orient_current} for a range of perturbation sizes, this illustrates the success of the optimizer when pushed near the limit. At the training angular scale $\sigma_{\mathrm{T},\mathrm{a}}>6^{\circ}$, adequate solutions can no longer be found.

Lastly, the current perturbation, although 10 percent of the base current, shows only marginal improvements. $Q$ improves for each stellarator, with the largest improvement seen for LPQH. Despite these small improvements, the more notable effect is the lower stochastic $\Jb$ value for LPQA, shown in the bottom panel of Figure~\ref{fig:jf_orient_current} for a range of perturbation sizes. Since the currents enter the stochastic objective linearly via Biot--Savart, one can show that the stochastic squared flux reduces to the deterministic one with a larger effective weight. However, the problem remains nonconvex in the coil shape, and the final solution has no guarantee of outperforming the deterministic solution.

\subsection{LPQH Stochastic AL Method}
\label{sec:auglag}

\begin{table}[tb]
    \centering
    \caption{Numerical metrics for the deterministic (Det AL) and 50S AL LPQH optimizations. The 50S AL optimization
    reduces the mean perturbed-field error while
    substantially increasing the robustness ratio $Q$. The perturbations
    match those used for the standard optimizations in
    Table~\ref{tab:stdnumvals_qh}.}
    \label{tab:auglag_numvals}

    \setlength{\tabcolsep}{12pt}
    \begin{tabular}{@{}lcc@{}}
        \toprule
        Metric & Det AL & 50S AL \\
        \midrule
        $\Jb\;(\units)$   & $\mathbf{6.6\times10^{-7}}$ & $2.6\times10^{-6}$ \\
        $\Joos\;(\units)$ & $2.0\times10^{-5}$ & $\mathbf{1.4\times10^{-5}}$  \\
        $\Jnorm$          & $\mathbf{2.2\times10^{-3}}$ & $4.6\times10^{-3}$  \\
        $Q$               & $0.0325$ & $\mathbf{0.187}$              \\
        \bottomrule
    \end{tabular}
\end{table}

\begin{figure}
\centering
    \includegraphics[width=0.40\textwidth]{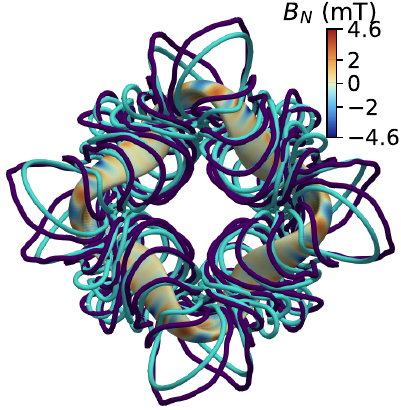}
    \caption{LPQH deterministic AL and 50S AL coil solutions overlaid on the plasma boundary. Dark purple curves are the deterministic solution; cyan curves are the stochastic solution. Displacements between the two families are noticeably larger than for the corresponding L-BFGS-B solutions in the top panel of Figure~\ref{fig:qh_geom}.}
    \label{fig:auglag_coils}
\end{figure}

Here we use the AL method as described in detail by Gil \textit{et al.}~\cite{gil_efficient_2026}, with all the optimization methodology from Sec.~\ref{sec:optimization model} carried over to this solver. We compare the deterministic AL method against the same algorithm and thresholds with a stochastic flux objective, as in Sec.~\ref{sec:pos_results}. We use the LPQH stellarator due to the difficulty of optimization and tuning for that configuration. We also change some of the desired thresholds compared with the LPQH runs from Sec.~\ref{sec:lpqh-results} and therefore these AL runs should not be directly compared. The goal here is to compare deterministic AL and stochastic AL.

Comparing the two AL solutions in Table~\ref{tab:auglag_numvals}, we see that the OOS evaluation, $\Joos=2.0\times10^{-5}\ \units$, is 30 times larger than its nominal flux $\Jb=6.6\times10^{-7}\ \units$, giving $Q=0.0325$ and indicating large sensitivity to perturbations.

The stochastic AL solution behaves differently. Its nominal objective is $\Jb=2.6\times10^{-6}\ \units$, while $\Joos=1.4\times10^{-5}\ \units$ is only about 5.4 times larger, giving $Q=0.187$. Relative to the standard stochastic solution, both quantities improve by an order of magnitude. The corresponding coil solutions are compared in Figure~\ref{fig:auglag_coils}.

\begin{figure*}
    \centering
    \includegraphics[width=0.47\textwidth]{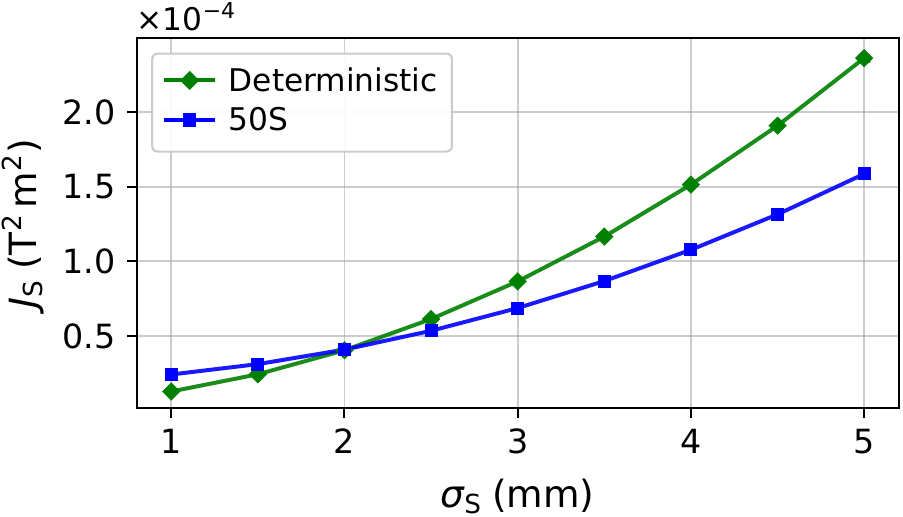}
    \includegraphics[width=0.47\textwidth]{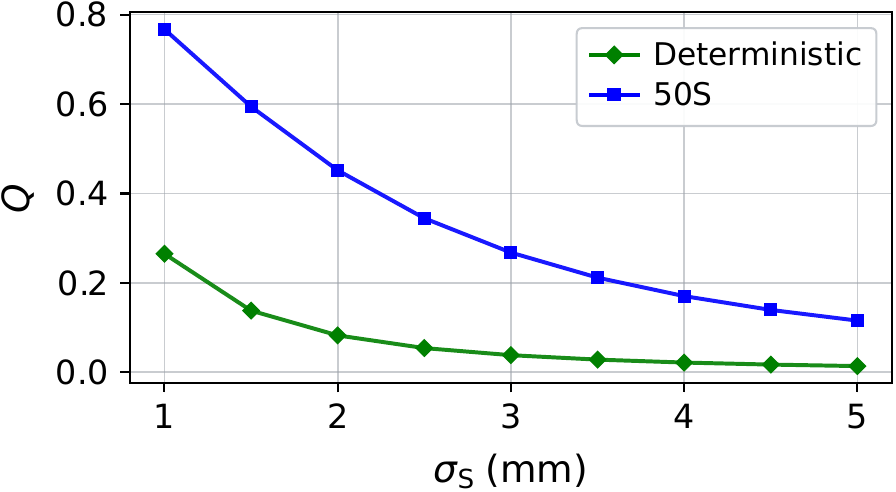}
    \includegraphics[width=0.47\textwidth]{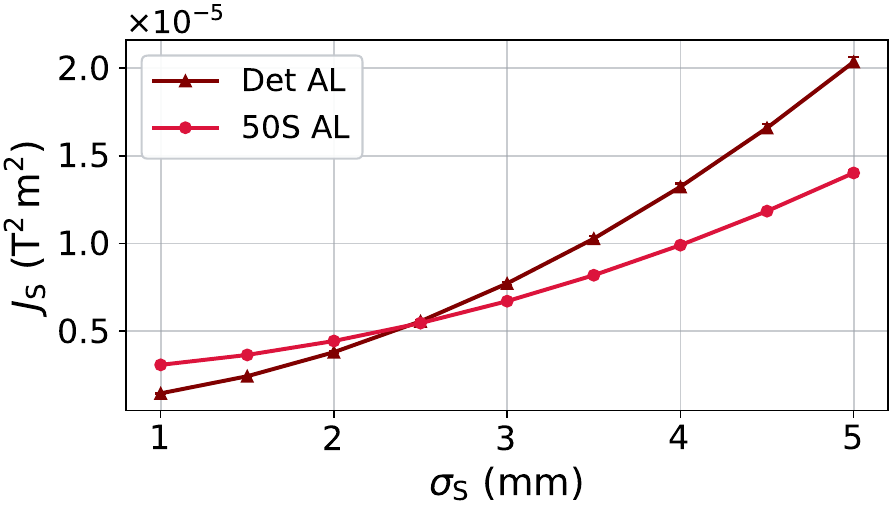}
    \includegraphics[width=0.47\textwidth]{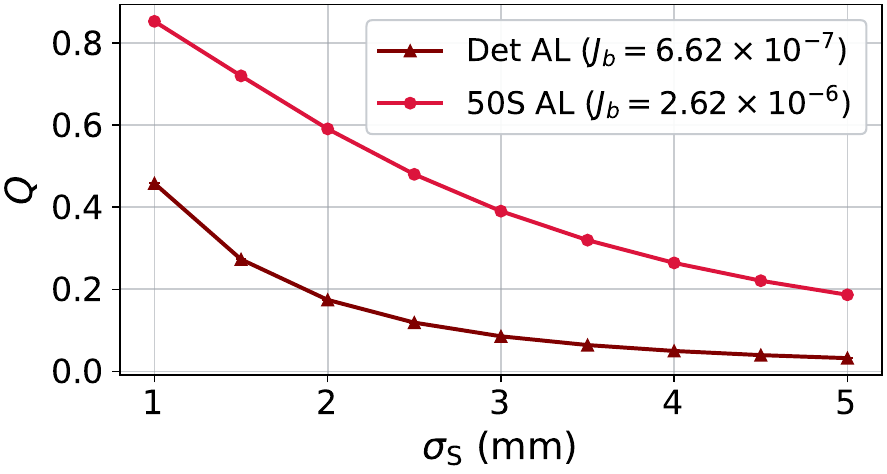}
    \caption{Direct out-of-sample evaluations for the standard and AL LPQH solutions. The top row shows mean $\Joos$ and $Q$ for the standard deterministic and 50S solutions over the OOS scale $\sigs$. The bottom row shows the corresponding AL results. Error bars give the standard error of the mean over $N_{\mathrm{S}}$ perturbation samples. }
    \label{fig:qh_tolerance}
\end{figure*}

\subsection{Effect of Sample Count}
\label{sec:samplecount}

The $N = 4$ results show that the benefit of the stochastic objective depends materially on sample count. Not only does $N = 4$ underperform $N = 50$, it fails to consistently improve on the deterministic baseline. Its out-of-sample flux exceeds the deterministic value over much of the LPQA and NCSX scans. At small $N$, the sample average poorly estimates the expectation, and the optimizer fits an unrepresentative draw of perturbations rather than the perturbation distribution itself. Sample count thus has a practical floor below which stochastic optimization is counterproductive rather than merely less effective.

\subsection{Discussion}
\label{sec:discussion}

Across the three configurations, stochastic optimization produces lower mean OOS flux than deterministic optimization, but the size of the change is configuration-dependent. The improvement is substantial for LPQH, clear for LPQA, and modest for NCSX. Because $Q=\Jb/\Joos$ can increase when nominal $\Jb$ increases, the absolute $\Joos$ values and the widths of the OOS distributions must be considered alongside $Q$. In the reported cases, the stochastic solutions also have lower absolute $\Joos$, supporting the interpretation that they occupy less perturbation-sensitive regions of the optimization landscape.

\subsubsection{Better objective values}
\label{sec:discussion-better-objectives}

At larger perturbation scales, the main stochastic benefit is lower $\Joos$. At some tighter scales, stochastic training also finds a lower nominal $\Jb$. This indicates that neighborhood averaging can sometimes guide the optimizer toward a minimum that is both deeper and less sensitive, as seen in Figures~\ref{fig:qa_stats}, \ref{fig:qh_stats}, and~\ref{fig:jf_orient_current}.

\subsubsection{Configuration dependence}
\label{sec:discussion-configuration}

LPQA has a comparatively well-conditioned stage-two landscape, and LPQH is the most sensitive case. NCSX shows wide variation in the final solution depending on the initialization, but the different final coil solutions exhibit very similar values of the squared flux. The benefit of stochastic training depends on the fixed plasma boundary, the initial coil family, and the optimizer. 

\subsubsection{Tolerance Gain}
  We quantify the mean perturbation size you can tolerate in the coils gained by stochastic training. In the training-$\sigma$ scan, where a separate model is trained at each $\sigma_T$, matching $J_S$ between the deterministic and 50S curves gives averaged gains of $0.09\,$mm (NCSX), $0.38\,$mm (LPQA), and $1.35\,$mm (LPQH) over $\sigma_T = 1$--$10\,$mm derived from figures \ref{fig:qa_stats}, \ref{fig:ncsx_stats}, \ref{fig:qh_stats}. Decoupling the evaluation size $\sigma_S$ from the training size, $\sigma_T = 5\,$mm, and comparing a single trained deterministic and 50S solution over $\sigma_S = 1$--$5\,$mm gives mean gains for LPQH of $0.38\,$mm for our standard solver and $0.22\,$mm for the AL solver as derived Figure \ref{fig:qh_tolerance}.

\begin{figure*}
\centering
    \includegraphics[width=0.9\textwidth]{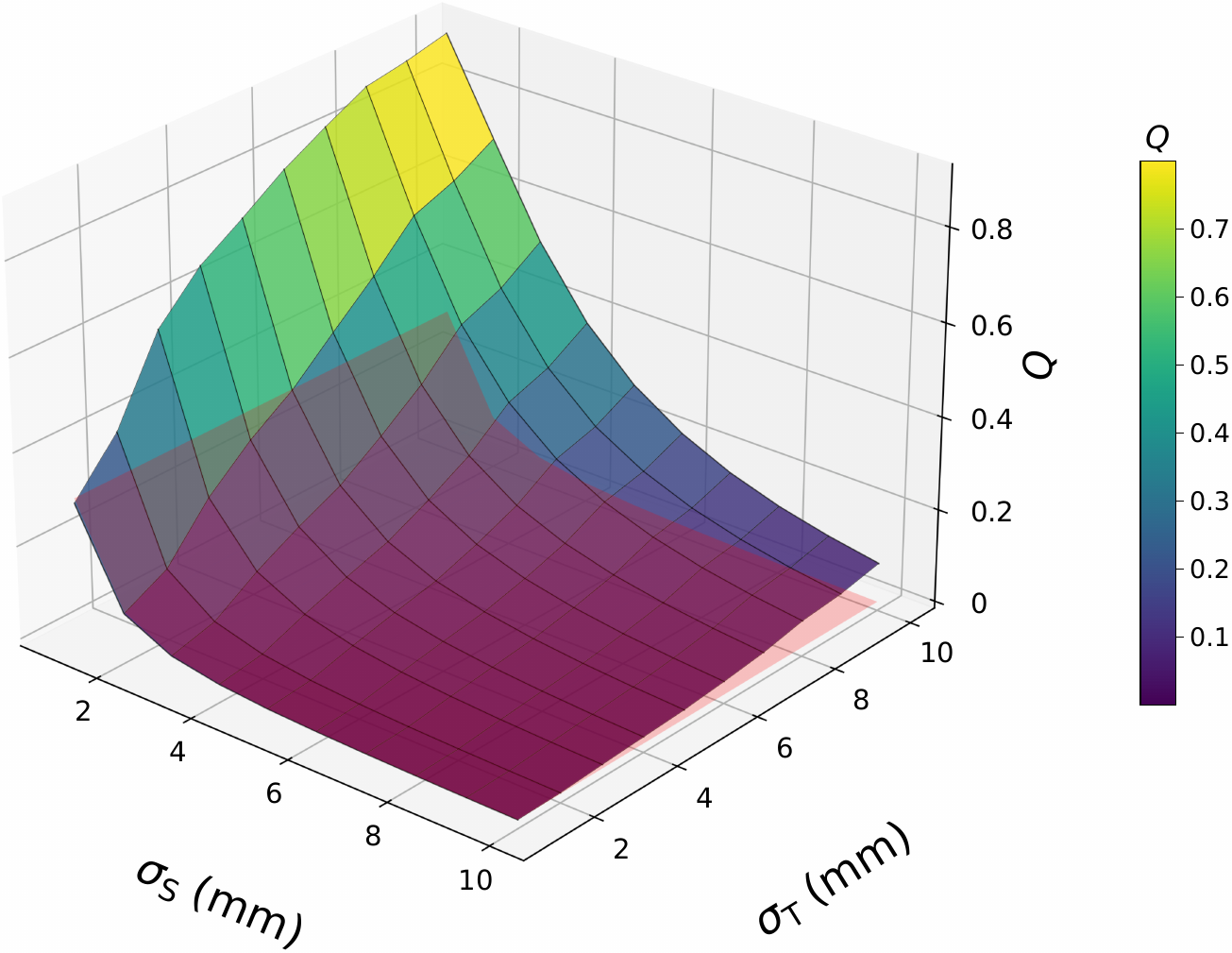}
    \caption{Quality number $Q$ for LPQH over the training scale $\sigt$ and OOS scale $\sigs$ for coil-shape perturbations. The flat overlay is the deterministic baseline, which has no training scale.}
    \label{fig:sigma_mesh}
\end{figure*}

\subsubsection{Joint training and evaluation scan}
\label{sec:sigma-mesh-discussion}

Figure~\ref{fig:sigma_mesh} varies $\sigt$ and $\sigs$ independently from 1\,mm to 1\,cm, with ten values on each axis. Each surface point therefore represents a separately trained stochastic solution evaluated against a specified OOS error distribution. The deterministic overlay provides the comparison at the same $\sigs$. The main takeaway is that robustness depends jointly on the distribution used during optimization and the distribution used during evaluation. Training with the largest available $\sigt$ does not always provide the most realistic coil solutions, so the intended error environment should be chosen before selecting the training distribution.

\subsubsection{OOS scans as an evaluation tool}
\label{sec:oos-scan-tolerance}

Figure~\ref{fig:qh_tolerance} holds each optimized coil set fixed and varies only the OOS scale $\sigs$. It therefore answers a different question from the training scan. The curves estimate expected squared-flux performance for one existing design under a family of Gaussian error distributions. Interestingly, we find that maximal improvements in $Q$ occur when $\sigma_S$ is small but $\sigma_T$ is large, i.e. stochastic optimization over a distribution of large perturbations tends to substantially improve the robustness with respect to small perturbations. 

\subsubsection{Recommended design workflow}
\label{sec:recommended-workflow}

A reasonable workflow is to scan $\sigt$ and evaluate candidate solutions at the intended $\sigs$. Select a training value by balancing nominal $\Jb$, mean $\Joos$, engineering constraints, and solver dependence. Then run a denser $\sigs$ scan on the chosen fixed solution. 

\subsubsection{Computational cost}
\label{sec:discussion-cost}

The primary cost of stochastic optimization is evaluating $N$ perturbed configurations per gradient step. The 50S runs distribute these evaluations across MPI processes. Biot--Savart evaluations parallelize well, but stochastic optimization still requires substantially more work than one deterministic objective evaluation. 

\section{Conclusion and Future Work}
\label{sec:conclusion}

\subsection{Conclusion}
\label{sec:summary}

Stellarator coil design is a nonconvex optimization problem, and deterministic solutions can be sensitive to the starting point and to geometric or current errors. Across the three fixed plasma boundaries studied here, stochastic stage-two optimization lowers $J_S$, the mean OOS flux. For the symmetric-start solutions, $Q$ increases by approximately 4.5\% for NCSX and 729\% for LPQH. The NCSX benefit is very modest, while the LPQH result shows a much larger trade between nominal fidelity and perturbation tolerance.

The AL calculations show that solver choice can materially change the robustness obtained from stochastic training. These findings extend earlier single-stage stochastic studies~\cite{wechsung_single-stage_2022,gil_stochastic_2026} by isolating coil robustness for fixed plasma boundaries. They also reinforce the need to report the error distribution, field normalization, solver, initial condition, and both nominal and OOS metrics when making manufacturability claims.

\subsection{Future Work}
\label{sec:future-work}

Several natural extensions of this work follow from the results presented here. The most immediate is checking coupled perturbations, e.g. manufacturing and current errors, as only independent simulations were run to ensure the results were beneficial to the optimization.
A second is a systematic Hessian analysis of the optimized coil solutions, which would provide a rigorous but \textit{local} geometric characterization of the minima located by the stochastic and deterministic optimizers. While the perturbation-size scans presented here are consistent with the stochastic method locating flatter, broader minima, a direct computation of the local curvature of the loss surface at each solution would confirm whether the empirical robustness improvement corresponds to a quantifiably smaller curvature along perturbation directions, strengthening the theoretical grounding of the stochastic approach and providing a criterion for comparing solution quality beyond $Q$. 

A third direction concerns the quality of the initial guess provided to the stage-two optimizer. The perturbed initial guess analysis demonstrates that the basin of attraction reached is sensitive to the starting configuration, particularly for NCSX and LPQH. One promising avenue is the use of winding-surface solutions for filamentary coil initialization~\cite{yu2024quasi,fu2026flexible,liang2026optimized}. This approach constrains the initial coil configurations to a physically meaningful region of the optimization space, potentially placing the optimizer closer to high-quality basins from the outset and reducing sensitivity of the final solution to the arbitrary choice of starting point. Combined with the stochastic objective and AL method, a well-chosen coil initialization could allow the optimizer to begin in a more favorable region of the landscape and converge to solutions that are simultaneously better and more robust.

Further open directions include the investigation of stochastic stage-one optimization, which would allow plasma boundary robustness to be considered jointly with coil robustness earlier in the design pipeline. Better optimization algorithms, including global algorithms, may further change the possibilities for more robust final solutions.

\begin{table*}[tb]
    \centering
    \caption{Coil-optimization parameters, thresholds, and weights for the
    LPQA, NCSX, LPQH, and LPQH AL configurations.
    The AL formulation enforces the geometric constraints
    directly, so the corresponding penalty weights are not applicable. One coil in the initialization was fixed at 100 kA of current, so the magnetic fields averaged on-axis differ in each device. The authors make no claim that these parameter choices are optimal; they serve as fixed parameters for comparing deterministic and stochastic optimization.}
    \label{tab:usrvars}

    \small
    \renewcommand{\arraystretch}{1.08}
    \begin{tabular*}{\textwidth}{@{\extracolsep{\fill}}llcccc@{}}
        \toprule
        \multirow{2}{*}{Category}
        & \multirow{2}{*}{Parameter}
        & \multicolumn{4}{c}{Configuration} \\
        \cmidrule(lr){3-6}
        & & LPQA & NCSX & LPQH & LPQH AL \\
        \midrule

        \multirow{7}{*}{Geometry}
        & \# of base coils, $N_b$                  & 4   & 3    & 5    & 5   \\
        & Field periods, $n_{\mathrm{fp}}$    & 2   & 3    & 4    & 4   \\
        & Total coils, $2n_{\mathrm{fp}}N_b$    & 16  & 18   & 40   & 40  \\
        & Fourier order, $M$ & 24  & 24   & 16   & 16  \\
        & Initial coil major radius, $R_0$ ($\mathrm{m}$)  & 1.0 & 1.3  & 1.0  & 1.0 \\
        & Initial coil minor radius, $R_1$ ($\mathrm{m}$)  & 0.5 & 0.85 & 0.5  & 0.5 \\
        & Mean initial reference field, $\langle B\rangle_{\mathrm{axis}}$ ($\mathrm{T}$)
            & 0.324 & 0.256
            & 0.812 & 0.812 \\
        & Plasma minor radius, (m) & 0.1683 & 0.3226 & 0.1411 & 0.1411 \\
        & Plasma aspect ratio & 6.0 & 4.4 & 7.0 & 7.0 \\
        \midrule

        \multirow{6}{*}{Length and distance}
        & Length threshold, $L_{\mathrm{max}}$ ($\mathrm{m}$)
            & 15.5 & 15.15 & 12.31 & 16.2 \\
        & Length weight, $w_L$ ($\mathrm{m}^{-2}$)
            & $5.0\times10^{-5}$
            & $8.0\times10^{-4}$
            & $1.0\times10^{-3}$
            & N/A \\
        & Coil-to-coil threshold, $d_\mathrm{CC}$ ($\mathrm{m}$)
            & 0.1 & 0.2 & 0.084 & 0.1 \\
        & Coil-to-coil weight, $w_{\mathrm{CC}}$ ($\mathrm{m}^{-2}$)
            & 100 & 100 & 100 & N/A \\
        & Coil-to-surface threshold, $d_\mathrm{CS}$ ($\mathrm{m}$)
            & 0.1 & 0.15 & 0.1 & 0.1 \\
        & Coil-to-surface weight, $w_{\mathrm{CS}}$ ($\mathrm{m}^{-2}$)
            & $4.0\times10^{-4}$
            & $4.0\times10^{-4}$
            & 10
            & N/A \\
        \midrule

        \multirow{6}{*}{Curvature and other}
        & Curvature threshold, $\kappa_{\mathrm{max}}$ ($\mathrm{m}^{-1}$)
            & 5.0 & 5.0 & 13.0 & 16 \\
        & Curvature weight, $w_{\kappa}$ ($\mathrm{m}^{2}$)
            & $1.0\times10^{-6}$
            & $1.0\times10^{-6}$
            & $1.0\times10^{-6}$
            & N/A \\
        & MSC threshold, $c_{\mathrm{max}}$ ($\mathrm{m}^{-2}$)
            & 5.0 & 5.0 & 1.3 & 16 \\
        & MSC weight, $w_{\mathrm{MSC}}$ ($\mathrm{m}^{4}$)
            & $1.0\times10^{-6}$
            & $1.0\times10^{-6}$
            & $1.5\times10^{-8}$
            & N/A \\
        & Arclength weight, $w_A$
            & 0.01 & 0.01 & 0.007 & N/A \\
        & Link weight, $w_{\mathrm{LK}}$
            & 1 & 1 & 100 & N/A \\
        \bottomrule
    \end{tabular*}
\end{table*}

    \begin{table*}[tb]
    \centering
    \small
    \renewcommand{\arraystretch}{1.08}
    \begin{tabular*}{\textwidth}{@{\extracolsep{\fill}}ll cc cc cc cc@{}}
        \toprule
        \multirow{2}{*}{Category} & \multirow{2}{*}{Parameter}
        & \multicolumn{2}{c}{LPQA} & \multicolumn{2}{c}{NCSX} & \multicolumn{2}{c}{LPQH} & \multicolumn{2}{c}{LPQH AL} \\
        \cmidrule(lr){3-4} \cmidrule(lr){5-6} \cmidrule(lr){7-8} \cmidrule(lr){9-10}
        & & Det & 50S & Det & 50S & Det & 50S & Det & 50S \\
        \midrule
        Geometry
        & Total length, $L_{\mathrm{tot}}$ ($\mathrm{m}$)
            & 15.6 & 15.7 & 15.3 & 15.4 & 12.4 & 12.4 & 16.31 & 15.26 \\
        \midrule
        \multirow{2}{*}{Curvature}
        & $\sum \kappa_{\mathrm{max}}$ ($\mathrm{m}^{-1}$)
            & 16.3 & 18.0 & 18.1 & 17.5 & 45.1 & 46.3 & 39.49 & 41.49 \\
        & $\sum \mathrm{MSC}$ ($\mathrm{m}^{-2}$)
            & 20.4 & 19.8 & 20.4 & 19.5 & 78.2 & 86.6 & 80.03 & 74.94 \\
        \midrule
        \multirow{2}{*}{Distance}
        & Min coil-to-coil, $d_{\mathrm{CC}}$ ($\mathrm{m}$)
            & 0.10 & 0.10 & 0.20 & 0.20 & 0.08 & 0.08 & 0.093 & 0.086 \\
        & Min coil-to-surface, $d_{\mathrm{CS}}$ ($\mathrm{m}$)
            & 0.274 & 0.305
            & 0.188 & 0.191
            & 0.100 & 0.119
            & 0.111 & 0.111 \\
        \midrule
        Field
        & Average optimized reference field, $B_{\mathrm{axis}}$ ($\mathrm{T}$)
            & 0.31 & 0.31
            & 0.24 & 0.24
            & 0.67 & 0.67
            & 0.22 &  0.18 \\
            
        \bottomrule
    \end{tabular*}
    \caption{Achieved coil-geometry and field metrics for the deterministic (Det) and 50-sample stochastic (50S) solutions across LPQA, NCSX, LPQH, and the LPQH augmented-Lagrangian (AL) benchmark. Length, curvature, and mean-squared curvature are summed over the base coil set per field period.}
    \label{tab:coilmetrics}
\end{table*}

\section*{Acknowledgments}
\label{sec:acknowledgments}
The authors acknowledge NYU High Performance Computing for resources, services, and staff expertise, the SIMSOPT development team, and the Simons Foundation under award 560651. We thank Pedro Gil for useful discussions. We thank the United States Department of Veterans Affairs and funding provided by the Chapter 31 Veteran Readiness and Employment program. We thank the Department of Applied Physics at New York University's Tandon School of Engineering for their recognition by awarding the 2026 Erich Kunhardt Prize for Innovative Senior Project for the project that preceded this one, whose findings are the basis for the manufacturing perturbation analysis section. 

\section{AI Use Statement}
\label{sec:AI_use}
Large language models developed by Anthropic PBC and OpenAI PBC were used throughout this work for large data analysis, proofreading, and for generating figures and tables from run records. All code and text were reviewed and edited by the authors, who take full responsibility for the content of this article.


\section{Data Availability Statement}
\label{sec:data_availability}
The code used to generate the numerical results is openly available at \url{https://doi.org/10.5281/zenodo.22822236}. The datasets and figures produced by this analysis, including the 35 additional plots referenced in the main text, are openly available at \url{https://doi.org/10.5281/zenodo.22822156}.

\appendix
\section{Coil Weights, Thresholds and Parameters}
\label{sec:appendix}

The coil weights, thresholds, and parameters are listed in Table~\ref{tab:usrvars}. For LPQA and NCSX, a range of weights and thresholds were tested, and the values shown produced the best balance of coil engineering metrics and $J_\text{norm}$. We make no claim that the chosen weights and thresholds are globally optimal.

We used the following plasma boundaries, which are available open-source in SIMSOPT: 
\begin{itemize}\item \texttt{input.LandremanPaul2021\_QA}, \item \texttt{input.20210406-01-002-nfp4\_QH\_000\_000240}, \item  \texttt{input.NCSX\_c09r00\_halfTeslaTF}. 
\end{itemize}
The Wiedman \textit{et al.}~\cite{wiedman_coil_2024,wiedman_data_2023} QH boundary corresponds to \texttt{input.LandremanPaul2021\_QH}. Its Fourier coefficients are not identical to those in the LPQH file used here, but we do not attempt to make a direct comparison to Wiedman \textit{et al.}; instead, we use it to estimate the parameters at which one can obtain a high-quality solution for this stellarator.

\renewcommand{\bibsection}{\section*{References}\phantomsection\label{sec:references}}
\bibliographystyle{unsrtnat}
\bibliography{references}

\end{document}